\documentstyle[epsf,psfig]{mn}
\def\xmm{{\em XMM-Newton}~}
\def\mch{M$\rm^{c}$Hardy\,}
\def\etal{\rm et al.~\rm}

\def\ecs{ergs cm$^{-2}$ s$^{-1}$~}
\def\ltsim{\mathrel{\hbox{\rlap{\hbox{\lower4pt\hbox{$\sim$}}}\hbox{$<$}}}}
\def\gtsim{\mathrel{\hbox{\rlap{\hbox{\lower4pt\hbox{$\sim$}}}\hbox{$>$}}}}

\begin{document}

\title[X-ray Variability of NGC~4051]
{\bf Combined Long and Short Timescale X-ray Variability of NGC~4051 with
{\it RXTE} and {\em XMM-Newton} }

\author[M$\rm^{c}$Hardy, I.M., \etal]
{ I.M. M$\rm^{c}$Hardy$^{1}$, I.E. Papadakis$^{2}$, P. Uttley $^{1}$,
M.J. Page$^{3}$ and K.O. Mason,$^{3}$ \\
$^{1}$ Department of Physics and Astronomy, The University, Southampton 
SO17 1BJ\\
$^{2}$ Physics Department, University of Crete,
Heraklion, Crete, Greece\\
$^{3}$ Mullard Space Science Laboratory, University College London, 
Holmbury St Mary, Dorking RH5 6NT
}
\maketitle
 
\begin{abstract}
We present a comprehensive examination of the X-ray variability of the
narrow line Seyfert 1 (NLS1) galaxy NGC~4051, one of the most variable AGN in
the sky. We combine over 6.5 years of frequent monitoring observations
by {\it RXTE} with a $>100$~ks continuous observation by {\em XMM-Newton} and so present an
overall 2-10 keV powerspectral density (PSD) covering an unprecedent
frequency range of over 6.5 decades from $<10^{-8}$ to $>10^{-2}$
Hz.  The combined {\it RXTE} and {\em XMM-Newton} PSD is a very good match to the PSD of
the galactic black hole binary system (GBH) Cyg~X-1 when in a `high',
rather than `low', state providing the first definite confirmation 
of an AGN in a `high' state.
We also find that a bending powerlaw, rather
than a sharply broken powerlaw, besides being more physical, is a much
better description of the high state PSD of Cyg~X-1 and is also a better
description of the PSD of NGC~4051.

At low frequencies the PSD of NGC~4051 has a slope of -1.1 bending, at
a frequency $\nu_{B}=8^{+4}_{-3}\times10^{-4}$Hz, to a slope of
$\alpha_{H} \sim -2$. Although $\nu_{B}$ does not depend on photon
energy, $\alpha_{H}$ is steeper at lower energies.  If $\nu_{B}$ scales
with mass, we imply a black hole mass of $3^{+2}_{-1} \times 10^{5}
M_{\odot}$ in NGC~4051, which is consistent with the recently reported
reverberation value of $5^{+6}_{-3} \times 10^{5} M_{\odot}$.
Hence NGC~4051 is emitting at $\sim30\% \rm~ L_{Edd}$.

NGC~4051 follows the same rms-flux relationship as GBHs, consistent with 
higher Fourier frequencies being associated with smaller radii.

From the cross-powerspectra and cross-correlation functions between
{\em XMM-Newton} lightcurves in different energy bands, we note that the higher
energy photons lag the lower energy ones. We also note that the lag is
greater for variations of longer Fourier period and increases with the
energy separation of the bands.  Variations in different wavebands are
very coherent at long Fourier periods but the coherence decreases at
shorter periods and as the energy separation between bands increases.
This behaviour is again similar to that of GBHs, and of MCG-6-30-15,
and suggests a radial distribution of frequencies and photon energies
with higher energies and higher frequencies being associated with
smaller radii.

Combining our observations with observations from the literature we
find it is not possible to fit all AGN to the same linear scaling
of break timescale with black hole mass. However
broad line AGN are consistent with a linear scaling of break
timescale with mass from Cyg~X-1 in its low state and
NLS1 galaxies scale better with Cyg~X-1 in its high state. We suggest
that the relationship between black hole mass and break timescale
is a function of at least one other underlying parameter which may be
accretion rate or black hole spin or both.

\end{abstract}
 
 
\section{INTRODUCTION}
 
Early observations with EXOSAT showed that, on short timescales (day
to minutes) AGN X-ray variability was quasi-scale invariant, with
power spectral densities (PSDs) being reasonably described by simple
powerlaws or perhaps slightly curving functions (Lawrence \etal 1987;
\mch and Czerny 1987).  However examination of archival data (\mch
1988) showed that, on longer timescales ($\gtsim$weeks) the scale
invariance of at least one AGN, NGC~5506, broke down and its PSD
flattened below a bend, or `knee' frequency of $\sim 10^{-6}$Hz.  The
combined long and short timescale PSD of NGC~5506 resembled that of
galactic black hole X-ray binary systems (GBHs) such as Cyg~X-1 (eg
Nowak \etal 1999) which, when in a `low' state, has a `knee' frequency
of $\sim 0.2$Hz, below which the PSD flattens to a slope of zero. If
the knee frequencies scale with mass, then a reasonable black hole
mass in NGC~5506 of $\sim$few $\times 10^{6}M_{\circ}$ was
implied. Thus the possibility arose of determining AGN black hole
masses from observations of long timescale X-ray variability.

Again using a variety of archival data, Papadakis and \mch (1995)
demonstrated that the PSD of the brighter AGN, NGC~4151, also flattened
at low frequencies ($<10^{-7}$Hz) but long term X-ray lightcurves
remained of poor quality until the launch of the Rossi X-ray Timing
Explorer ({\it RXTE}; Swank 1998) in December 1995.  {\it RXTE} can sample on
sub-daily timescales and so, since shortly after launch we, and other
groups, have been monitoring a small sample of AGN in order to produce
high quality long timescale PSDs. One of our main targets is NGC~4051
and results from the first 1.5 years of observation (\mch
\etal 1998) indicated that the slope at low frequencies was probably
flatter than that at high frequencies but neither the break frequency,
or the slopes, were particularly well defined. 
Edelson and Nandra (1999), Pounds
\etal (2001) and Markowitz \etal (2003) have all now demonstrated PSD
flattening in other AGN at low frequencies, using {\it RXTE} data.

Although {\it RXTE} provides much better long term X-ray lightcurves than
any previous observatory, the lightcurves are still not continously
sampled, or even completely uniformly sampled. Thus the `raw' or
`dirty' PSDs will be contaminated by artefacts which depend on the
window sampling pattern. In order to overcome such problems we have
developed a simulation-based modelling process (Uttley, M$\rm^{c}$Hardy
and Papadakis 2002)
which can determine not only the value of any break frequencies, or PSD
slopes, but can also derive errors for these parameters, and also
quantify the goodness of fit of any model. Uttley \etal (2002) applied
this method to the first 3 years of observations of four AGN. They
found evidence for bends, or breaks, in the PSDs in three AGN,
NGC~5506, NGC~3516 and MCG-6-30-15, although not in NGC~5548.

GBHs are found in a variety of `states' (eg McClintock and Remillard 2003; 
Cui \etal
1997a), the most common of which are the so called low flux, hard
spectrum (or `low') state, and the high flux, soft spectrum (or
`high') state. In the standard Comptonisation models for the
production of X-ray emission, different physical regimes apply to low
and high states, with a higher seed photon flux in the high state,
leading to a cooler Comptonising corona (eg Zdziarski \etal 2002).  If
we are to attempt physical modelling of AGN we therefore need to know
which state they are in. For GBHs, the PSDs of the high (eg Cui
\etal 1997b, Churazov \etal 2001) and low (eg Nowak \etal 1999;
Revnivtsev, Gilfanov and Churazov 2000) states are different, thereby
providing us with a means of discrimination.

In Cyg~X-1 in the low state, the PSD slope above the `knee' (refered
to above, at $\sim 0.2$Hz) is $\sim -1$. However at $\sim 3$Hz there
is a second, higher frequency, break above which the PSD slope is
$\sim-2$. In the high state only one break has been detected so far in
the PSD of Cyg~X-1.  The break frequency is $\sim10-20$Hz and, above
the break, the PSD slope is steeper than -2. Below the break the PSD
has a slope of -1 for at least 4 decades of frequency (Reig \etal
2002).

For AGN, there are two cases (Akn~564, Papadakis \etal 2002 and
NGC~3783, Markowitz \etal 2003) where two breaks have been found,
indicating a possible similarity to low state systems. However in most other
AGN only one break has been found and, although the slopes below
the break are not that well defined (eg Uttley \etal 2002; Markowitz
\etal 2003), in general they are closer to -1 than to zero. However it
is not clear whether these breaks correspond to the high frequency
break in low state systems or the single break in high state systems.

If we assume that timescales scale approximately with mass, then to
properly compare AGN with GBHs we require AGN PSDs which cover
timescales from $\sim$few years to tens of seconds.  {\it RXTE} monitoring
observations principally sample timescales from $\sim$few years to
$\sim$day but {\em XMM-Newton} can provide continous observations of up to almost 2
days with sampling on $\sim$second, or shorter, timescales.
NGC~4051 is the AGN which is best observed on
long timescales by {\it RXTE} and is also one of the best observed on short
timescale with {\em XMM-Newton}.  Here we present a combined PSD analysis of these two
datasets.

In Section~\ref{sec:obs} we describe the observations and the data
analysis procedures. In Section~\ref{sec:psd} we discuss the combined
{\it RXTE} and {\em XMM-Newton} PSD, covering more than 6 decades in
frequency, and compare it to that of GBHs, particularly Cyg~X-1, in
different states.  As further diagnostics of the Comptonisation
process we discuss the high frequency {\em XMM-Newton} data in more
detail, and note that a similar analysis of a long {\em XMM-Newton}
observation of MCG-6-30-15 has been carried out by Vaughan, Fabian and
Nandra (2003), producing broadly similar results.  In
Section~\ref{sec:psd} we note a variation of PSD slope with photon
energy at high frequencies.  In Section~\ref{sec:rms} we determine the
rms-flux relation (Uttley and \mch 2001) from the {\em XMM-Newton}
data.  We discuss the relationship between variations in different
wavebands both in terms of the cross-correlation function
(Section~\ref{sec:crosscor}) and using cross-spectrum phase spectrum
analysis (Section~\ref{sec:crosspectrum}).
In Section~\ref{sec:coherence} we discuss the coherence of the
variations between different wavebands, ie whether the same variations
occur in all bands or whether there are independent contributions in
different wavebands.  In Section~\ref{sec:summary} we summarise the
main observational results from this paper. Finally, in
Section~\ref{sec:implications}, we compare NGC~4051 with high and low
state GBHs and discuss possible physical models for the X-ray emission
region.  We also examine the relationship between PSD break timescale
and black hole mass in AGN, noting that the relationship is probably different
for broad and narrow line Seyfert 1 galaxies, perhaps because of a
difference in mass accretion rate or spin.

\section{OBSERVATIONS AND DATA REDUCTION}
\label{sec:obs}

\subsection{{\it RXTE}}
\begin{figure*}
\psfig{figure=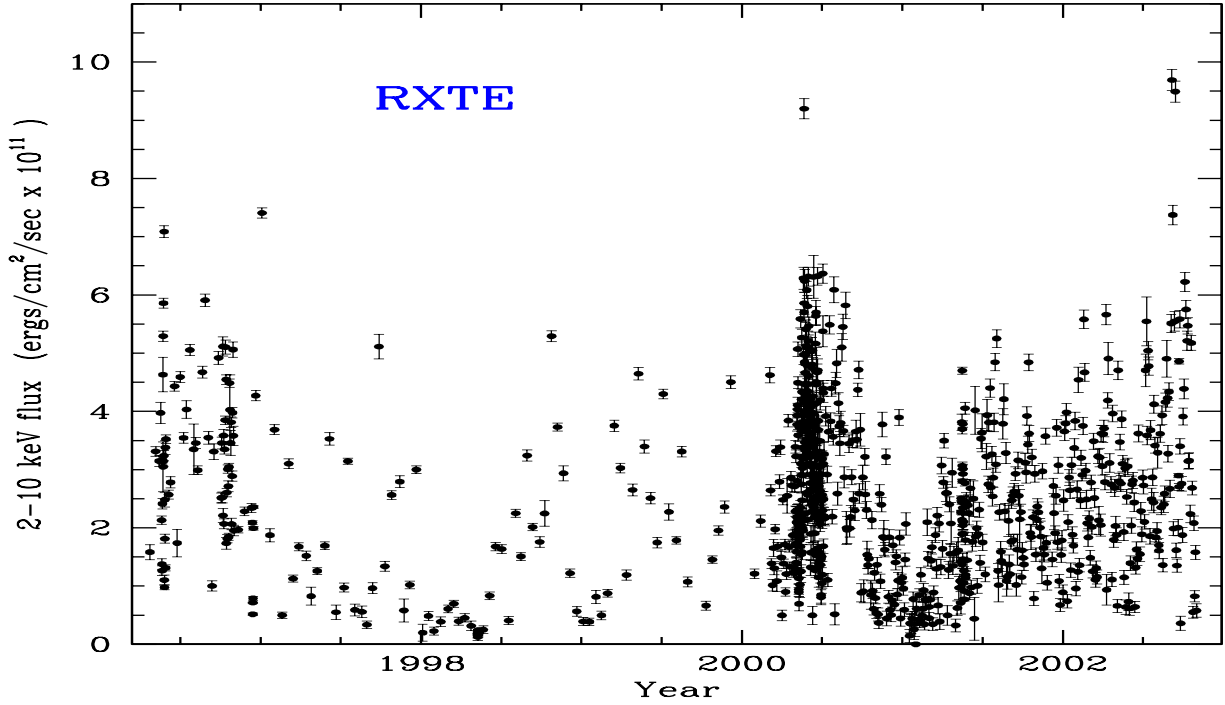,width=6.8in,height=4in,angle=0}
\caption{{\it RXTE} Long Term 2-10 keV lightcurve of NGC~4051. Each data point
represents an observation of $\sim1$~ks.}
\label{fig:4051xtelc}
\end{figure*}

Here we present (Fig.~\ref{fig:4051xtelc}) all of the short
($\sim$1~ks duration) monitoring observations of NGC~4051 which have
been carried out with the proportional counter array (PCA, Zhang \etal
1993) on {\it RXTE} from 1996 until 2002 inclusive.  As the gain of
the PCA, and the number of PCUs, has varied throughout the lifetime of
the instrument, we derive here a lightcurve in flux units rather than
in counts so that all observations may be used together.  
This `flux' lightcurve is used in all subsequent analysis.

During periods when the gain, and number of PCUs did not change, the
flux and count rate lightcurves are identical, indicating no
significant contribution from variable absorption to the variability
above 2 keV which is seen by {\it RXTE}.  (We also note that, in their
detailed study of the X-ray spectral variability of NGC~4051 with {\it
RXTE}, Lamer \etal (2003a) find no evidence that the time averaged spectral
variations are caused by variable absorption. Also from {\it RXTE}
observations Taylor \etal (2003) find that the X-ray spectral
variability of NGC~4051 is well explained by pivoting of the X-ray
spectrum about a high energy ($\sim100$ keV) and do not require any
variations of absorbing column.)

Changes in luminosity can alter the ionisation state of warm gas in
the line of sight to AGN such as NGC4051 (eg \mch \etal 1995) and so
can affect the transparency of the gas.  Such changes do not affect
significantly the spectrum above 2 keV and are mainly restricted to
lines from carbon, oxygen, nitrogen and neon below 2 keV. However, as
we see from the \xmm reflection grating spectrum (Ogle \etal 2003)
which was taken at the same time as the \xmm imaging observations
described below, the total absorption below 2 keV in those lines in
NGC4051 is only a few percent of the total continuum flux. Therefore
variations in that few percent would be small and could not account
for the large flux variations seen in Fig.~\ref{fig:4051xmmlc}.
Thus the flux variations described in this
paper represent variations in the continuum source and not in any
absorbing gas.

The PCA consists of 5 Xenon-filled proportional counter units (PCUs)
sensitive to X-rays with energies between 2 and 60 keV. The maximum
effective area of the PCA is 6500 cm$^{2}$.  For each observation we
extracted the Xenon 1 (top) layer data from all PCUs that were
switched on during the observation as layer 1 provides the highest S/N
for photons in the energy range 2-20 keV where the flux from AGN is
strongest.  We used FTOOLS v4.2 for the reduction of the PCA data and
extracted source spectra according to the standard method outlined in
Lamer \etal (2003b). We used standard `good time' data selection
criteria, i.e. target elevation $>10^{\circ}$, pointing offset
$<0.01^{\circ}$, time since SAA passage $>30$min and standard
threshold for electron contamination.  We calculated the model
background in the PCA with the tool PCABACKEST v2.1 using the L7 model
for faint sources.  PCA response matrices were calculated individually
for each observation using PCARSP V2.37, taking into account temporal
variations of the detector gain and the changing number of detectors
used.  Fluxes in the 2-10 keV band were then determined using XSPEC,
fitting a simple powerlaw with variable slope but with absorption
fixed at the Galactic level of $1.3 \times 10^{20}$ cm$^{-2}$ (Elvis,
Lockman and Wilkes 1988; Stark \etal 1992; \mch \etal 1995).
The errors in the flux are scaled directly from the observed
errors in the measured count rate.

As can be seen from Fig.~\ref{fig:4051xtelc}, a variety of
sampling patterns were undertaken. In the first four AOs, prior to
2000, we used a quasi-logarithmic sampling pattern, covering all
timescales, but rather economically. In order to improve
the S/N on our resulting PSDs, and to improve our understanding of
other phenomena such as `low' states (eg Guainazzi \etal 1998;
Uttley \etal 1999), and of the relationship between the X-ray and
optical flux variations (Peterson \etal 2000; Shemmer \etal 2003),
we then increased our coverage. From 2000 onwards, our minimum
observation frequency has been once every two days. In addition,
to sample properly the higher frequencies, we observed every 6 hours
for 2 months in 2000.

\subsection{{\em XMM-Newton}}

\begin{figure*}
\psfig{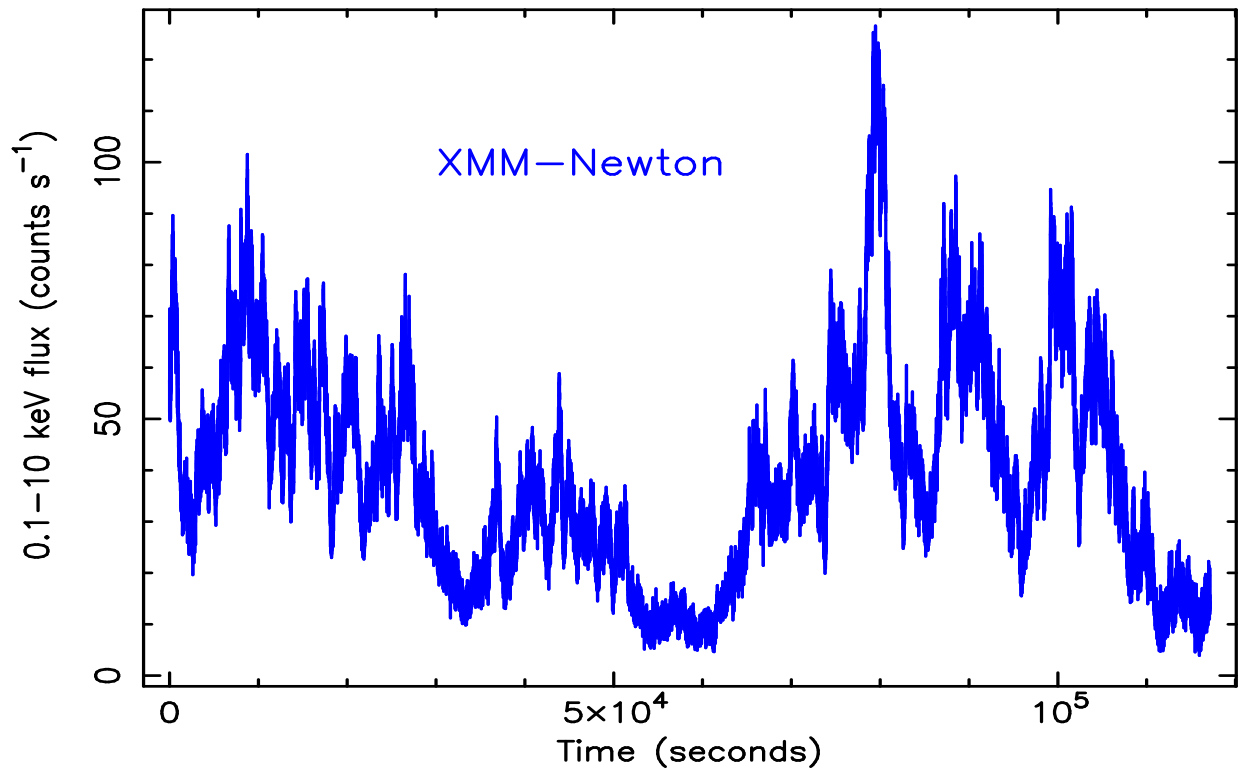}
\caption{XMM-Newton background-subtracted lightcurve of NGC~4051 in 
the 0.1-10 keV energy band, with 5s time bins, combining data from
both PN and MOS CCDs.}
\label{fig:4051xmmlc}
\end{figure*}

NGC~4051 was observed for a complete {\em XMM-Newton} orbit over the 16th and
17th May 2001. The EPIC PN and MOS2 cameras were operated in small
window modes with medium filters, while MOS1 was in timing mode with
the thin filter. The PN and MOS cameras took data continuously for 117
ks and 107 ks respectively.  Background levels were low in all 3
detectors for the first 103 ks, after which the background began to
flare due to increased levels of soft protons, affecting the
lightcurves significantly at high energies ($>$5 keV).  

The data were processed and lightcurves were produced using the
{\em XMM-Newton} Standard Analysis System (SAS) version 5.2. The PN
source+background lightcurve was obtained from a 42 arcsecond radius
circle centred on the source, and the background lightcurve was taken
from two similar sized source-free regions on the same chip. The MOS2
source+background lightcurve was taken from a 45 arcsecond circle
centred on the NGC~4051 and the background lightcurve was obtained from
source-free regions of the inner parts of the outer chips. The MOS1
source+background lightcurve was obtained from a 52-pixel wide region
of the timing window centred on the source and the background
lightcurve was taken from a 26-pixel wide strip at the edge of the
timing window.

The PN detector in small window mode, and MOS1 in timing mode, have
negligible pileup at the typical flux of NGC~4051 and, apart from a
difference in S/N, the lightcurves are the same within the
errors. However MOS2, in small window mode, is slightly affected by
pileup.  An empirical correction for pileup, of up to 10\% at the
highest countrates, in the MOS2 lightcurve was therefore obtained by
comparison with the MOS 1 and PN lightcurves. After 
applying this time-averaged correction,
the MOS2 and PN lightcurves are the same within the errors.
The background subtracted and (for MOS2) pileup corrected lightcurves
were then summed to produce combined lightcurves of the highest
possible S/N.  The resulting full band (0.1-10 keV) lightcurve is
shown in Fig.~\ref{fig:4051xmmlc}.  As the full 0.1-10 keV lightcurve
is dominated by photons of low energy, whose background level is not
greatly affected by the proton flare referred to above, we show the
full 117~ks of the observation in Fig.~\ref{fig:4051xmmlc}. However,
for consistency, we restrict analysis, in all bands, to the first
102.5~ks of the observation, when the hard bands are not affected by
increased background.

The {\em XMM-Newton} lightcurves, which we binned at 5s resolution,
contain a small number ($\sim14$) of short gaps. With the exception of
one gap of 105s and another of 200s, all gaps have durations $\leq
35$s with most being $\leq 20$s. If left in, as zero counts, these
gaps introduce a very small amount of spurious high frequency power
into the PSDs and very slightly flatten it at frequencies
$<10^{-2}$Hz. We have therefore filled these gaps by linear
interpolation and the addition of noise characteristic of the
surrounding datapoints. As the gaps are so small, and have such a
small effect on the PSD (as one can see by inserting similar gaps
elsewhere into the {\em XMM-Newton} lightcurves), the difference
between linear interpolation and any other reasonable interpolation is
undetectable.  We therefore work in the rest of this paper with these
lightcurves, whose units are `counts s$^{-1}$', as their
continuous nature simplifies some of the later analysis.

\section{Powerspectral Analysis}
\label{sec:psd}

\subsection{Analysis Method}
\label{sec:analysis}

In order to model reliably the PSD of our data, taking into account
the distorting effects of gaps in the observed {\it RXTE} light
curves, together with red-noise leak effects, we use the Monte Carlo
simulation based {\sc psresp} method of Uttley et al. (2002).  {\sc
psresp} is based on the earlier `response method' of Done et
al. (1992), which simulates red-noise lightcurves assuming a given
PSD model and the observed sampling pattern, and compares the
resulting model PSD
with the observed PSD.  {\sc psresp} additionally allows PSDs measured
from a number of lightcurves covering a range of time-scales to be
combined, as is required here.  Furthermore, the {\sc psresp} method
uses the distribution of best fitting $\chi^{2}$ of the simulated PSDs
to estimate reliably the confidence for the null hypothesis that the
assumed PSD model is correct.  We refer the reader to Section~4 of
Uttley et al. (2002) for a full discussion of the {\sc psresp} method.
In the context of that discussion, we note the following features of
our {\sc psresp} analysis of the {\it XMM-Newton} and {\it RXTE} PSDs:

\begin{enumerate}
\item For ease of computation we made two {\it RXTE} lightcurves, 
to cover long (years - month) and medium (month - day) timescales.
The long timescale lightcurve is made from the entire {\it RXTE}
monitoring lightcurve, binned up to 2-week resolution. The medium
timescale lightcurve is made from only the two-month-long section of
6~hourly sampling and is binned to exactly 6 hour resolution.  The
binned lightcurves are very complete. Apart from a 200 day period at
the end of 1999/beginning of 2000 where there are 9 observations
instead of 14, there are no gaps. During the 64 day period of the
6~hourly sampling there are 247, rather than 256,
observations. However during periods when there were 3 rather than 4
observations per day, the 3 observations were fairly evenly spaced. In
both cases the few empty bins are filled by linear interpolation from
adjacent bins.  From the {\it XMM-Newton} data we made a lightcurve
binned to 500~s resolution to make a high-frequency PSD, which is also
included in the simulations.  Since it is computationally prohibitive
to simulate lightcurves at the full 5~s resolution of our {\it
XMM-Newton} lightcurves, we insert directly into {\sc psresp} the very
high frequency (VHF: $10^{-3}$--$2\times10^{-2}$~Hz) part of the PSD
which is made from the full resolution lightcurve.
This PSD is binned logarithmically
into bins of frequency width $1.5\nu$, with errors determined in the
conventional way, using the method of Papadakis \& Lawrence (1993). The
errors are well-determined due to the many cycles sampled at each
frequency. As the observations are continuously sampled to frequencies
above those included in the VHF PSD, where Poisson noise is dominant,
aliasing is not a problem. Similarly, as the lightcurve samples
frequencies well below those included red noise leak is
not a problem.

\item For a given PSD model we simulated evenly sampled 
lightcurves, using the method of Timmer and Konig (1995), with time
resolutions of 6 hours, 36 minutes and 500~s for the long and medium
timescale {\it RXTE} lightcurves and the {\it XMM-Newton} lightcurve
respectively.  These lightcurves were then resampled
to match the observed sampling pattern. The simulated
{\it RXTE} lightcurves were then rebinned to the same resolution
(2 weeks and 6 hours) as the observed lightcurves.

\item Both simulated and observed PSDs
are binned up in logarithmically spaced frequency intervals of width
$\nu\rightarrow1.5\nu$.  Poisson noise
levels are added to the simulated PSDs
together with the `aliased' contribution expected due to
variations shorter than the time resolution of the simulated light
curves (see Uttley et al. 2002 for details).

\item For each set of PSD model parameters tested we
simulate $N=300$ realisations for each of the three lightcurves
included in the fit. For each of these three lightcurves 
we actually simulate one long lightcurve of length $300$ times
the length of the observed lightcurve, and then split it into sections,
thereby reducing red noise leakage
problems.  We therefore assume no power on timescales longer than $300
\times$ the length of each of the contributing lightcurves, 
ie about 2000 years for the longest lightcurve.  We use $M=3000$
random combinations of the simulated PSDs (plus realisations of the
VHF PSD) to compare with the model average PSDs, in order to determine
the distribution of $\chi^{2}$ expected for the given model (see
Uttley et al. 2002, Section 4.2).  Comparison with the observed
$\chi^{2}$ yields a probability $P$ that the model is acceptable.
Following the practice of Uttley et al. 2002, we define `absolute'
90\% errors on all model parameters, based on the parameter values
which bracket the region of the parameter grid which is acceptable at
$P>0.1$.
\end{enumerate}

\subsection{PSD of the NGC~4051 {\it RXTE} Data}
\label{sec:xtepsd}

As the long timescale PSD covers a large frequency range, it provides
a first indication of whether NGC~4051 might be more like a low or
high state system. We therefore modelled the
{\it RXTE} data on its own, treating it as two separate lightcurves of
6 hour and 2 week resolution, as described above.

\begin{figure}
\psfig{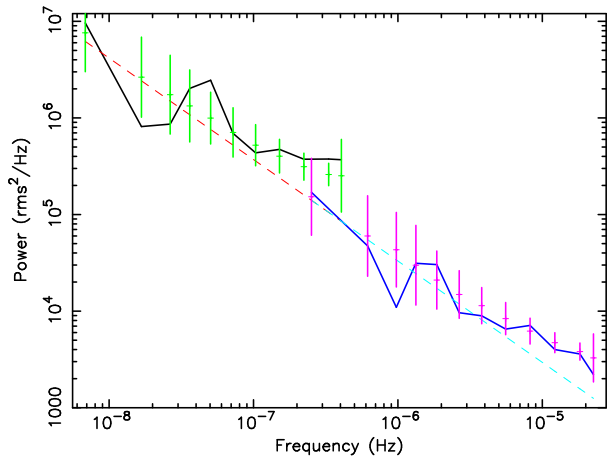}
\caption{PSD of NGC~4051 using {\it RXTE} data only.
The underlying single powerlaw model is the straight dashed line. The
distorted model is given by the points with errorbars and the observed
data is given by the two continuous solid lines.  The best-fit powerlaw slope
is $-1.05^{+0.15}_{-0.20}$ and the fit probability is 93\% }
\label{fig:xtepsd}
\end{figure}

We first fitted the simplest model, ie a single unbroken
powerlaw. That model turned out to be an excellent fit to the {\it RXTE}
data (Fig.~\ref{fig:xtepsd}).  The powerlaw slope, over the
frequency range $7 \times 10^{-9}$ to $2 \times 10^{-5}$Hz was
$-1.05^{+0.15}_{-0.20}$ and the fit probability was 0.93. (Throughout
this paper we quote 90\% confidence errors unless stated otherwise.)
There was therefore no justification for fitting
more complex models to that frequency range.

The large frequency range, fitted by a slope of -1.05, indicates that
NGC~4051 may very well be a high state system. Therefore 
before making more complex fits, including the {\em XMM-Newton} data, we re-examine
high state observations of Cyg~X-1 in order to obtain a good PSD
template with which we may compare the PSD of NGC~4051.

\section{The High State PSD of Cygnus X-1}

\subsection{Calculation of the Cyg~X-1 High State PSD}

\begin{figure*}
\psfig{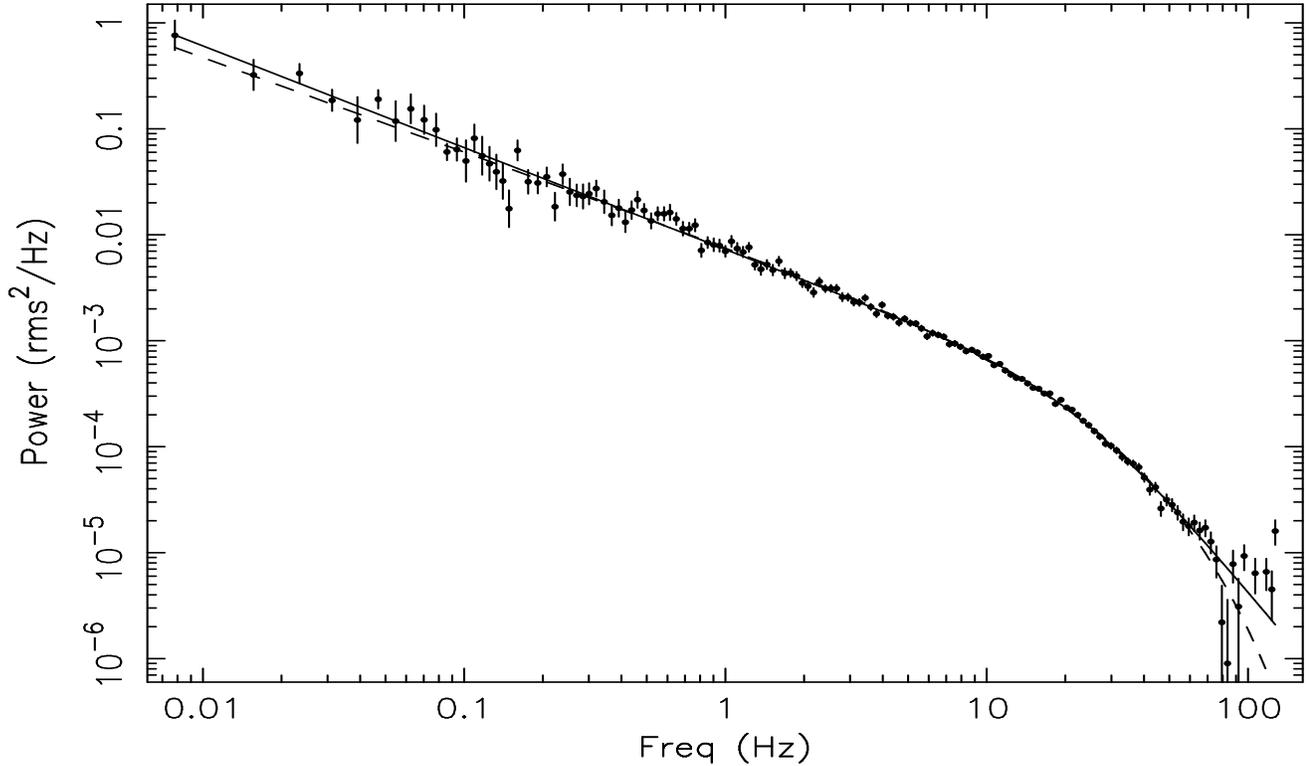}
\caption{PSD of Cyg~X-1 in the high state in the 2-13 keV range. 
The dashed line is a
powerlaw with an exponential cut-off and the solid line is a bending
powerlaw, as described in the text. The upper frequency of the fits
was 80Hz, but the fits have been extrapolated to higher frequencies.
Although the bending powerlaw is a marginally better fit, both
models are formally good fits.}
\label{fig:cygpsd}
\end{figure*}

\begin{figure}
\psfig{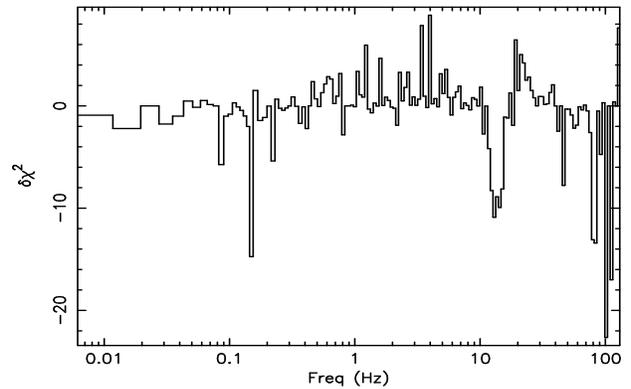}
\caption{$\delta \chi^{2}$ residuals for the best fitting sharply broken powerlaw
(see text for parameters) to the PSD of Cyg~X-1 shown in Fig.~\ref{fig:cygpsd},
fitting over the same frequency range. Note the large systematic deviations
near the break frequency (13.9 Hz).
}
\label{fig:cygresid}
\end{figure}

Cyg~X-1 is the archetypal GBH against whose PSD the PSDs of AGN are
usually compared. We should therefore determine first which model best
describes the high state PSD of Cyg~X-1 and then test that model
against our observations of NGC~4051. A high state PSD of Cyg~X-1 has
been published by Revnivtsev \etal (2000) from {\it RXTE}
proposal P10512 (see also Cui \etal 1997b). Revnivtsev \etal find that,
below 15-20 Hz, the PSD has a slope of -1 and at higher frequencies
the slope steepens to $\sim-2$. However they do not attempt to fit a
single continuous model to the whole frequency range ($\sim 10^{-3}$
to $\sim$ few$\times 10^{2}$Hz) covered by the {\it RXTE} observations. We
have therefore recalculated the high state PSD. We used PCA standard
binned mode data from a single-orbit {\it RXTE} observation of Cyg~X-1
in the high state (Observation ID 10512-01-09-01, observed 1996 June
18) and produced lightcurves in the 2-5 keV, 5-8 keV, 8-13 and 2-13
keV bands. (Note, sensitivity is low below 2.5 keV).  The lightcurves
are continuous and the resultant PSDs are not distorted noticeably by
the window function and so can be considered to be a good
representation of the true underlying PSDs.  The final PSDs were
obtained by averaging the PSDs measured from 128~s segments of the
lightcurve and, when there is more than one frequency in a bin,
binning in logarithmic frequency intervals
$\nu\rightarrow1.05\nu$. We estimated standard errors in each
frequency bin using the spread in individual powers about the mean,
and applied the deadtime correction formula of Sunyaev
\& Revnivtsev (2000) to the Poisson noise level to obtain
Poisson-noise-subtracted PSDs.  The PSDs from this observation are
fairly typical of those obtained in the high state in this source (Cui
\etal 1997b).  The PSDs in various bands look rather similar and so we
show, in Fig.~\ref{fig:cygpsd}, just the broad band 2-13 keV PSD.

\subsection{Modelling of the Cyg~X-1 High State PSD}

Discontinuous broken powerlaws of the form
\[ P(\nu) \propto \nu^{\alpha_{L}} \mbox{ for $\nu < \nu_{B}$ and} \]  
\[ P(\nu) \propto \nu^{\alpha_{H}}  \mbox{ for $\nu > \nu_{B}$}  \] 

\noindent
are often used to describe PSDs. In the present case, such a fit to
the full 2-13 keV PSD gives $\alpha_{L}
=-1.025 \pm 0.013$, $\nu_{b} = 13.9 \pm 0.8$ and
$\alpha_{H} = -2.13 \pm 0.06 $. These values
are very similar to those derived by Revnivtsev \etal ( $\alpha_{low}
\sim-1$, $\alpha_{high} \sim-2.1$, $\nu_{b} \sim 10-20$).  However the
fit is poor, particularly around the break frequency
($\chi^{2}=242/134 \, dof$, ie P $< 0.1\%$, see Fig.~\ref{fig:cygresid}).  
It is clear that a
smooth curve rather than a sharp break describes Fig.~\ref{fig:cygpsd}
better. We have therefore fitted two smoothly curving functions to the
PSD. The dashed line in Fig.~\ref{fig:cygpsd} is the best fit of a
powerlaw with an exponential cut-off. The solid line is the best fit
of a function which smoothly changes from one powerlaw to
another. Below we give the form of that function which can accomodate
any number of joined powerlaws where the powerlaw at low frequencies
bends smoothly to become a steeper powerlaw at higher frequencies.
\[ P(\nu) = A \nu^{-\alpha_{1}} \,\Pi_{i=1}^{N}\, \left( 1 + 
\left(\frac{\nu}{\nu_{b_{i}}} \right)^{(\alpha_{i+1} - \alpha_{i})}
\right)^{-1}\, \]

\begin{table*}
\centering
\caption {\bf BENDING POWERLAW FITS TO CYG X-1}
\begin{tabular}{cccccc}
Energy Band & Normalisation (A) & $\alpha_{L}$ & $\alpha_{H}$ & $\nu_{B}$ & $\chi^{2}/dof$\\
(KeV)       &       &       &              &  (Hz) & \\
& & & & & \\   
2-5 &$5.44 \times 10^{-3}$& $-0.939\pm 0.026$ & $-2.72\pm0.20$ & $20.1\pm2.2$ & 124/134\\
5-8 &$1.00 \times 10^{-2}$& $-0.945\pm 0.026$ & $-2.87\pm0.26$ & $21.9\pm2.3$ & 135/134\\
8-13&$1.19\times 10^{-3}$& $-0.943\pm 0.024$ & $-3.18\pm0.35$ & $23.7\pm2.2$ & 136/134\\
\end{tabular}
\label{tab:cygall}
\end{table*}

\begin{table*}
\centering
\caption {\bf BENDING POWERLAW FITS TO CYG X-1 WITH FIXED $\alpha_{L}=-0.94$}
\begin{tabular}{ccccc}
Energy Band & Normalisation  (A)& $\alpha_{H}$ & $\nu_{B}$ & $\chi^{2}/dof$\\
(KeV)       &        &              &  (Hz) & \\
& & & & \\   
2-5 &$5.44\times 10^{-3}$&  $-2.72\pm0.14$ & $20.2\pm1.1$ & 124/135\\
5-8 &$1.00\times 10^{-2}$&  $-2.84\pm0.19$ & $21.5\pm1.2$ & 135/135\\
8-13&$1.19\times 10^{-3}$&  $-3.16\pm0.28$ & $23.5\pm1.4$ & 136/135\\
\end{tabular}
\label{tab:cyglow}
\end{table*}

\noindent
In the present case we use it to describe just one break in the PSD
($N=$ number of breaks), and we refer to it as a bending powerlaw fit.
The parameter, $A$, is the power at 1~Hz, assuming $\nu_{B} >> 1$Hz, or 
the power extrapolated from the lowest frequency powerlaw if there is
a break below, or near to, 1~Hz.  $A$ is refered to in a number of
following Tables as the normalisation. We have not used quite as much
data as Revnivtsev \etal and have not attempted, as they did, to
improve on the deadtime correction of Sunyaev \& Revnivtsev. We
therefore only fit models to the PSD up to 80Hz. However we plot
extrapolations of the fits to higher frequencies. Although the bending
powerlaw is a slightly better fit to the data ($\chi^{2}$=129/134 dof,
P=61\%), the bending powerlaw and exponential cut-off
($\chi^{2}$=140/135 dof, P=35\%) are both formally good descriptions
of the data, from $<10^{-2}$Hz to 80Hz. For the bending powerlaw fit,
the low frequency slope $\alpha_{low}, =-0.96 \pm 0.02$, the break
frequency $\nu_{b}, = 22.9 \pm 1.5$Hz and the high frequency slope,
$\alpha_{high}, = -2.99 \pm 0.14 $. For the exponential cut-off fit
the low frequency slope $\alpha_{low}, =-0.89 \pm 0.02$ and the break
frequency $\nu_{b}, = 23.7 \pm 1.1$Hz.  We note that although the
details of the fits vary, both the exponential and bending powerlaw
fits agree on the value of $\nu_{b}$.

Revnivtsev \etal who calculate the high state PSD of Cyg~X-1 up to
$>$200 Hz, favour a powerlaw description of the highest frequency part
of the PSD above the break. As a bending powerlaw is also the best 
fit to the PSDs which we have calculated, we therefore chose that 
model as our standard parameterisation of the high state PSD of 
\mbox{Cyg~X-1}.

\subsection{Variation of Cyg~X-1 High State PSD shape with energy}

The parameters of the fits of the bending powerlaw to the three energy
bands in Cyg~X-1 are given in Table~\ref{tab:cygall}. It can be seen
that $\alpha_{low}$ is the same in all fits to within the errors.
In order to reduce the
errors on the other parameters, further fits have therefore been made
with $\alpha_{low}$ fixed (at -0.94). The results of these fits are
given in Table~\ref{tab:cyglow}. From that table we can see that the
high frequency slope, $\alpha_{H}$, becomes slightly steeper and
$\nu_{B}$ increases slightly at higher energies although we caution
that $\nu_{B}$ is towards the upper edge of the measurable frequency
range and so $\alpha_{H}$ is not well constrained.

As $\alpha_{H}$ and $\nu_{B}$ are correlated in any fitting process,
we have performed separate fits, additionally fixing either
$\alpha_{H}$ (Table~\ref{tab:cyglowhigh}) or $\nu_{B}$
(Table~\ref{tab:cyglownu}).  In the first case we still note a slight
increase of $\nu_{B}$ with increasing energy and in the second case we
note a very slight steepening of $\alpha_{H}$ with increasing energy.
The changes between bands are small. For example in
Table~\ref{tab:cyglownu} we see that the range of $\alpha_{H}$ is only
0.2, which is about the same as the 90\% confidence error on each
measurement. More detailed observations are therefore needed to clarify
whether these changes are real. If they are real then we note that they are
in the opposite sense to that found for GBHs in the low state (eg Cyg~X-1,
Nowak \etal 1999) where the  PSD at high frequencies (2-90 Hz, ie above
the high frequency break)
becomes \underline{less} steep with increasing photon energy.

\section{Combined {\it RXTE} and {\em XMM-Newton} Powerspectrum of NGC~4051}
\label{sec:2-10kevpsd}

\subsection{PSD in 2-10 keV Band}
\begin{figure*}
\psfig{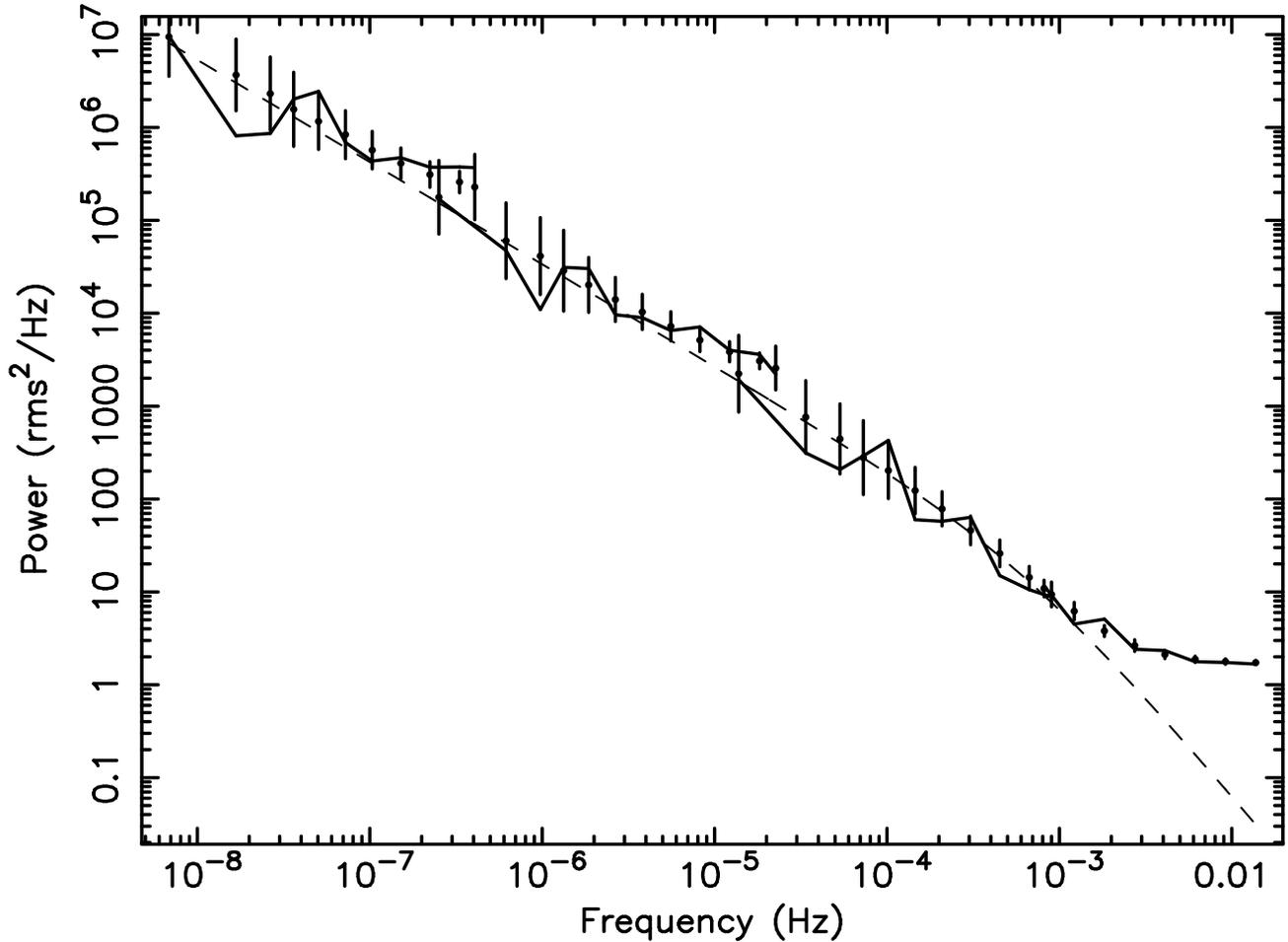}
\caption{ Combined {\it RXTE} and {\em XMM-Newton} (4-10 keV) PSD covering 6.5 decades of
frequency. The underlying model is the thin dashed line. The distorted
model is given by the points with errorbars and the observed data is
given by the four continuous solid lines. The two lower frequency
datasets ($7 \times 10^{-9}$ to $2 \times 10^{-5}$Hz) are from {\it RXTE}
and the two higher frequency datasets ($\sim 10^{-5}$ to $\sim
10^{-3}$Hz and $\sim 10^{-3}$ to $\sim 10^{-2}$Hz) are from {\em XMM-Newton}. See
the text for a more detailed description.  Note that the Poisson noise
has not been subtracted from the distorted model or the observed data
as the noise level is included as part of the fitting process.
Thus the PSD flattens at the highest frequencies.
}
\label{fig:xtexmmpsd}
\end{figure*}

In this section we calculate the combined long and short timescale PSD
of NGC~4051 by combining long timescale data from {\it RXTE} and short
timescale data from {\em XMM-Newton}.  As stated earlier
(Section~\ref{sec:obs}) the gain of the {\it RXTE} PCA detectors, and
the number of PCUs is not constant. We therefore cannot make useful
continuous lightcurves in units of counts s$^{-1}$ and so make
lightcurves in flux units, in the 2-10 keV band. The median photon
energy of that band is 5.1 keV. As the PSD shape may be a function of
photon energy (eg Nowak \etal 1999) we ensure that the median photon
energy of the {\em XMM-Newton} lightcurve which we combine with the
{\it RXTE} lightcurve is approximately the same. As {\em XMM-Newton}
has a much softer response than {\it RXTE}, the closest approximation
to the {\it RXTE} 2-10 keV band is the {\em XMM-Newton} 4-10 keV band,
which has a median energy of 4.9 keV.

The {\it RXTE} lightcurves are simulated in the same way as is described in
Sections~\ref{sec:analysis} and ~\ref{sec:xtepsd}, and the {\em XMM-Newton}
lightcurves are treated as is described in Section~\ref{sec:analysis}.
Overall we arrange the binning of the various lightcurves to ensure
that there is almost no overlap of the resultant parts of the final
PSD, which is shown in Fig.~\ref{fig:xtexmmpsd}.  This PSD is the best
PSD yet obtained for any AGN and is visually quite similar to the high
state PSD of Cyg~X-1 which is shown in Fig.~\ref{fig:cygpsd}. A quantitative
comparison of these PSDs is discussed below in Section~\ref{sec:modelfit}.

\begin{table}
\centering
\caption {\bf BENDING POWERLAW FITS TO CYG X-1 WITH FIXED $\alpha_{L}=-0.94$,
$\alpha_{H}=-3.0$}
\begin{tabular}{cccc}
Energy Band & Normalisation  (A)&  $\nu_{B}$ & $\chi^{2}/dof$\\
(KeV)       &                      &  (Hz) & \\
& & & \\   
2-5 &$5.27\times 10^{-3}$&  $21.3\pm0.9$ & 134/136\\
5-8 &$0.99\times 10^{-2}$&  $22.0\pm1.1$ & 137/136\\
8-13&$1.20\times 10^{-3}$&  $23.3\pm1.4$ & 137/136\\
\end{tabular}
\label{tab:cyglowhigh}
\end{table}

\begin{table}
\centering
\caption {\bf BENDING POWERLAW FITS TO CYG X-1 WITH FIXED $\alpha_{L}=-0.94$, $\nu_{B}=22.0$Hz}
\begin{tabular}{cccc}
Energy Band & Normalisation (A) &  $\alpha_{H}$ & $\chi^{2}/dof$\\
(KeV)       &                      &  (Hz) & \\
& & & \\   
2-5 &$5.27\times 10^{-3}$&  $-2.84\pm0.13$ & 132/136\\
5-8 &$0.99\times 10^{-2}$&  $-2.88\pm0.17$ & 135/136\\
8-13&$1.20\times 10^{-3}$&  $-3.05\pm0.25$ & 139/136\\
\end{tabular}
\label{tab:cyglownu}
\end{table}

As a test of whether there is any significant problem with red noise
leakage in the {\em XMM-Newton} VHF PSD, we have compared the results
of including the {\em XMM-Newton} data in two separate ways. Firstly
we treat the {\em XMM-Newton} data as a combination of a 500s binned
lightcurve and a VHF PSD, as described in
Section~\ref{sec:analysis}. Secondly we include the {\em XMM-Newton}
data simply as a PSD of the full 5s binned {\em XMM-Newton}
lightcurve, covering $\sim10^{-5}$ to $\sim2 \times 10^{-2}$Hz,
without any simulation of a 500s binned lightcurve. The combined {\it
RXTE} and {\em XMM-Newton} model fits are the same in both cases to
within the errors indicating that even without simulating the longer
timescales present in the {\em XMM-Newton} data, the raw {\em
XMM-Newton} PSD is untroubled by red noise leakage. Thus the VHF PSD
is completely free of such problems.  Below we discuss only the
results obtained by including, with the {\it RXTE} data, the {\em
XMM-Newton} data as a combination of a 500s binned lightcurve and a
VHF PSD.

\subsection{Model Fitting}
\label{sec:modelfit}
We first fitted a simple powerlaw to the PSD. The PSD slope is -1.25
but the fit is poor (P=2.7\%).

\begin{figure}
\psfig{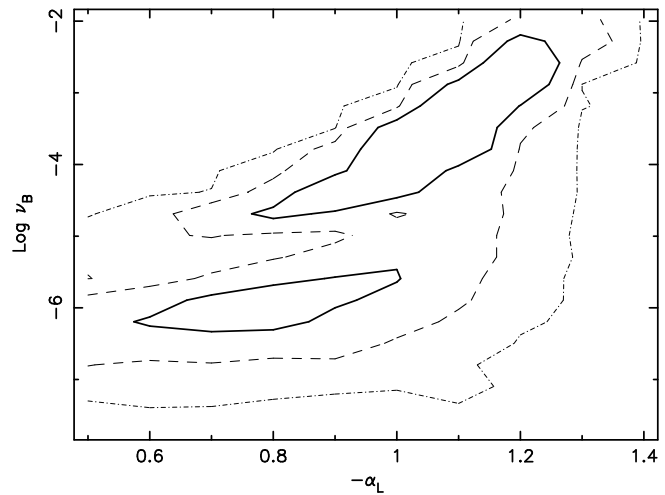}
\caption{68\%, 90\% and 99\% confidence contours for low frequency slope, $\alpha_{L}$,
and break frequency, $\nu_{B}$, for bending powerlaw fit to the combined
{\it RXTE} and {\em XMM-Newton} 4-10 keV PSD. Note here we plot $-\alpha_{L}$ to simplify
the labels.
}
\label{fig:xtexmm14}
\end{figure}

\begin{figure}
\psfig{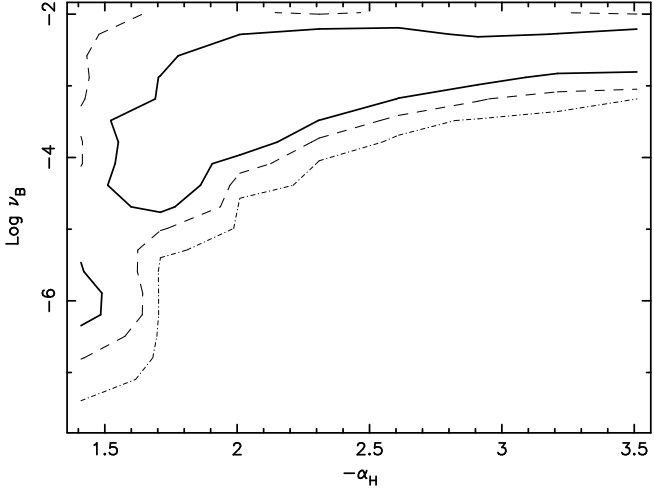}
\caption{68\%, 90\% and 99\% confidence contours for high frequency slope, $\alpha_{H}$,
and break frequency, $\nu_{B}$, for bending powerlaw fit to the combined
{\it RXTE} and {\em XMM-Newton} 4-10 keV PSD. Following Fig.~\ref{fig:xtexmm14} we plot
$-\alpha_{H}$.
}
\label{fig:xtexmm13}
\end{figure}

\begin{figure}
\psfig{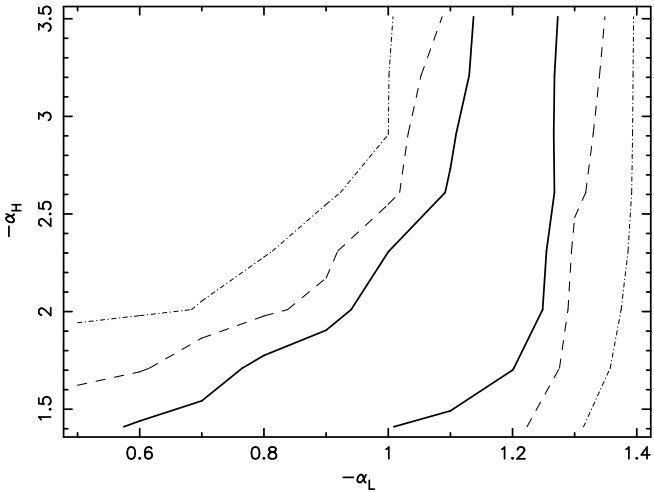}
\caption{68\%, 90\% and 99\% confidence contours for low and high frequency slopes,
$\alpha_{L}$ and $\alpha_{H}$ respectively for bending powerlaw fit to the combined
{\it RXTE} and {\em XMM-Newton} 4-10 keV PSD. Again we plot $-\alpha_{L}$
and $-\alpha_{H}$.
}
\label{fig:xtexmm34}
\end{figure}

We therefore next tried the bending powerlaw model which best fits the
Cyg~X-1 high state. This model is a very good fit (best fit
probability = 76\%). The best fitting low frequency slope,
$\alpha_{L}=-1.1$, the break frequency, $\nu_{B}=5\times10^{-4}$Hz and
the high frequency slope, $\alpha_{H}=-1.9$.  The data are quite
consistent with the values of $\alpha_{L}=-1$ and $\alpha_{H}=-2$
which Revnivtsev \etal (2000) use to describe the high state of
Cyg~X-1.
The best model fit and confidence contours for combinations of these
three parameters are given in figures~\ref{fig:xtexmmpsd},
\ref{fig:xtexmm14},
\ref{fig:xtexmm13} and  \ref{fig:xtexmm34}. 

We note that, although a very good fit is obtained with the above
parameters, it is not the only reasonable fit. A possible fit is
obtained with a much lower break frequency. However with a very low
break frequency (ie $\sim 10^{-6}$Hz)
we also require a high frequency slope which is
close to the best fitting low frequency slope
given above, and little difference between  $\alpha_{L}$ and
$\alpha_{H}$. This possible fit is another indication that most of
the frequency range of the combined {\it RXTE} and {\em XMM-Newton} PSD is explained
by a slope not far different from
$\sim-1.1$, although there may be some slight curvature at the
lowest observable frequencies.

Although a bending powerlaw is a much better fit to the Cyg~X-1 data,
for comparison we have fitted the
previously commonly used sharply breaking powerlaw to the combined {\em XMM-Newton} 
and {\it RXTE} data.
A good fit is obtained (P=82\%) with $\nu_{B}= 2.4 \times 10^{-4}$Hz
($90\%$ confidence limits of $2.4  \times 10^{-6}$ to $1.5  \times
10^{-3}$Hz), $\alpha_{L}=-1.1^{+0.4}_{-0.1}$ 
and $\alpha_{H}=-1.8^{+0.4}_{-0.8}$. The high and low frequency slopes 
are very similar to those found using the bending powerlaw model but 
$\nu_{B}$ is lower and is not particularly tightly constrained.

The overall conclusion is that a single powerlaw is not a good fit,
and two powerlaws are required, but the parameters of any bending or
breaking powerlaw are not well defined.

\subsection{ Variation of $Fvar$ with Energy for NGC~4051}

In Table~\ref{tab:fvar} we summarise the variability detectable in the
various {\em XMM-Newton} bands. In column 2 we give the average counts/sec (`Ave'),
and in column 3 we give what is known as $Fvar$, ie the square root of
the noise-subtracted variance, divided by the average count rate.
These values of $Fvar$ have been calculated from lightcurves which were
binned up to 100s in order to cut down the possibility of spurious
contributions to the noise (which must be subtracted from the
variance) from frequencies at which the PSD is completely dominated by
Poisson noise. 
We see that $Fvar$ decreases with photon energy, above 2 keV.
A similar decrease of variability with increasing photon energy
in AGN has been reported in the past by Green, \mch and Lehto (1993),
Nandra \etal 1997, 
Turner, George and Nandra (1998) and Markowitz and Edelson (2001).

The interpretation of this result is not unique. One possibility,
suggested by Green \etal, is
that there is a relatively constant component of hard spectrum.
That spectrum would not greatly affect the softer bands but would
dilute the variability in the harder bands. That interpretation is
consistent with the extremely hard component seen in a very low
flux state (Guainazzi \etal 1998; Uttley \etal 1999). However
Markowitz and Edelson (2001) claim, in their sample of AGN,
that the decrease in variability with increasing energy is
too strong to be explained only by the presence of a constant
hard component. Alternatively, if
in a Comptonisation scenario the bulk of the variability were provided 
by the seed photons, the greater number of scatterings required to
raise photons to higher energies should wash out high frequency
variability, but is contrary to the observations shown below
(Section~\ref{sec:highfreqpsd}). However Taylor \etal (2003)
show that the X-ray spectral variability of NGC~4051 is best
described by pivoting of the X-ray spectrum about a high energy
($\sim100$ keV), which automatically leads to a reduction in $Fvar$
with increasing energy (at least below the pivot energy).
In the Comptonising models of Zdziarski \etal (2002) spectral pivoting
is explained if the corona is responsible for dissipation of most
of the accretion energy and its luminosity remains constant whilst
the seed photon luminosity varies.

We note that $Fvar$ continues to rise down to energies below 1 keV.
We also note that the 2-10 keV flux measured by {\it RXTE} varies in
almost exactly the same way as the 0.1 keV flux seen by {\it EUVE}
(Uttley \etal 2000), except that the amplitude of variability of the
{\it EUVE} flux was slightly greater than that of the {\it RXTE} flux,
consistent with spectral pivoting extending to quite low energies, and
with the rise in $Fvar$ to low energies. Thus although the shape of
the spectrum below 2 keV in NGC~4051 is more complex than a simple
powerlaw (Salvi 2003 and Salvi \etal 2003; Uttley \etal 2003)
the large majority of the low energy flux appears to 
vary, which is different from Cyg~X-1 in the high state where there
is a strong soft component which does not vary greatly.

However in the very lowest energy bin (0.1-0.5 keV)
$Fvar$ actually decreases very slightly. Uttley \etal (2003) show
that there is a relatively constant component in NGC~4051 at low
as well as high (Taylor \etal 2003) energies and this component,
although small, may account for the slight decrease in $Fvar$.

\begin{table}
\centering
\caption {\bf Variation of $Fvar$ with Energy for NGC~4051}
\begin{tabular}{crc}
Energy Band & Ave & $Fvar$ \\
keV  & c s$^{-1}$    &  \\
&&  \\ 
     0.1-0.5 & 17.8 &   0.51 \\ 
       0.5-1 & 11.7 &   0.54 \\ 
         1-2 &  7.6 &   0.52 \\ 
         2-5 &  3.3 &   0.40 \\ 
        5-10 &  0.8 &   0.27 \\ 

\end{tabular}
\label{tab:fvar}
\end{table}

\subsection{Variation of High Frequency PSD Shape of NGC~4051 with Energy}
\label{sec:highfreqpsd}

\begin{figure}
\psfig{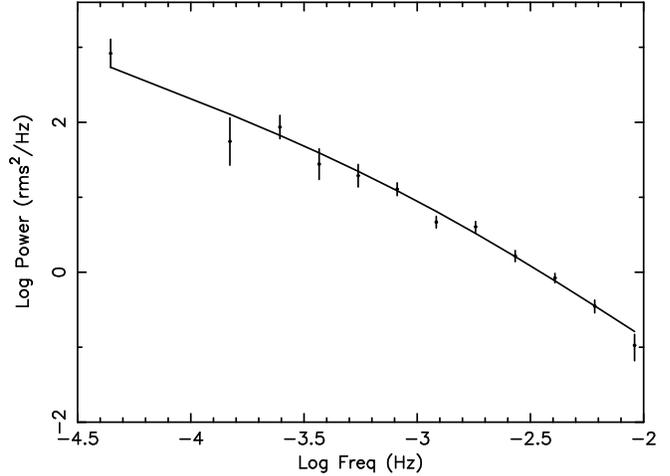}
\caption{PSD of NGC~4051 as measured by {\em XMM-Newton} in the 2-10 keV energy
band. The solid line is the best fit with $\alpha_{L}$ fixed at -1.1.
The best fit values are $\alpha_{H}=-2.03^{+0.4}_{-0.7}$ and
$\nu_{B}=8.3^{+19}_{-8.1}\times 10^{-4}$Hz.  The Poisson noise level has been
subtracted from the PSD.}
\label{fig:2-10psd}
\end{figure}

\begin{figure}
\psfig{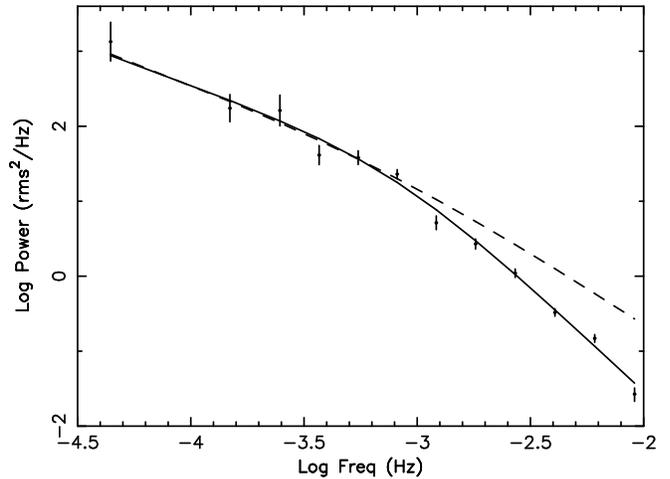}
\caption{PSD of NGC~4051 as measured by {\em XMM-Newton} in the 0.1-2
keV energy band. The solid line is the best fit with $\alpha_{L}$
fixed at -1.1.  The best fit values are
$\alpha_{H}=-2.83^{+0.2}_{-0.2}$ and $\nu_{B}=8.1^{+3.3}_{-2.9}\times
10^{-4}$Hz.  The Poisson noise level has been subtracted from the
PSD. The dashed line is the best 2-10 keV fit from
Fig.~\ref{fig:2-10psd} plotted with the same normalisation as that
of the 0.1-2 keV PSD.}
\label{fig:0.1-2psd}
\end{figure}

We have produced PSDs, using just the {\em XMM-Newton} data, in a
variety of spectral bands in order to investigate the spectral
dependence of the PSD shape.  We have already shown that the observed
{\em XMM-Newton} PSDs derived here represent very closely the true
PSDs. We can then remove the Poisson noise analytically and fit the
bending powerlaw model to the noise-subtracted PSDs. The fits,
allowing all parameters to be free, are given in
Table~\ref{tab:allfree}.  The fitting was carried out within the
plotting package, QDP. We note that although the fits to the various
{\em XMM-Newton} PSDs look quite good visually (eg
Figs.~\ref{fig:2-10psd} and \ref{fig:0.1-2psd}), the formal fits are
not particularly good for the lower energy bands where the errors on
the PSD datapoints are smaller. In those bands the main contribution
to the reduced $\chi^{2}$s comes from 2 or 3 datapoints with very
small errors. For any sort of smooth continuous function without sharp breaks,
or additional components such as QPOs,
these data points prevent us from obtaining 
a better fit to the 0.1-2 keV PSD with any other model.
Further observations would clarify the reality of the deviations from
a smooth function but for the moment we can only assume that
the deviations are a statistical fluctuation.

The errors on the parameters, allowing all to be free, are quite large
and so we next fix $\alpha_{L}$ at -1.1, the best fit value from the
joint {\it RXTE} and {\em XMM-Newton} fit. The resultant fits are
given in Table~\ref{tab:fixal}. From that Table we can see that
$\nu_{B}$ does not change with energy, to within the errors, but the
high frequency slope steadily flattens as we move to higher
energies. (The exception is the 5.0-10.0 keV band, but there are few
photons in that band and the errors on all parameters are very large.)

In addition to the five relatively narrow spectral bands listed at the
top of Tables~\ref{tab:allfree} and \ref{tab:fixal}, we calculate the
PSD in two broader bands, 0.1-2.0 and 2.0-10.0~keV, in order to
improve the statistical significance of the fits, which are given at
the bottom of Tables~\ref{tab:allfree} and \ref{tab:fixal}.  The
noise-subtracted PSDs themselves are shown in Figs.~\ref{fig:2-10psd}
and \ref{fig:0.1-2psd}.  We can see immediately that the two PSDs are
similar below $\sim10^{-3}$Hz but, at higher frequencies, the 0.1-2
keV PSD is much steeper.  However both PSDs have approximately the
same value of $\nu_{B}\sim8\times10^{-4}$Hz.

We can see, from Figs~\ref{fig:2-10psd} and \ref{fig:0.1-2psd} and
Table~\ref{tab:fixal}, that although the high frequency PSD slope is
steeper in the lower energy band, the normalisation of the PSD is
larger. If the normalisations and break frequencies were the same, the
steepening of the PSD would give rise to a lower value of $Fvar$ at
lower energies.  We therefore note that it is the higher normalisation
of the PSD at lower energies which accounts for the higher value of
$Fvar$.

\begin{table*}
\centering
\caption {\bf BENDING POWERLAW FITS TO {\em XMM-Newton} PSD OF NGC~4051}
\begin{tabular}{lcclrr}
Energy Band & Normalisation  (A)& $\alpha_{L}$ & \multicolumn{1}{c}{$\alpha_{H}$} & 
\multicolumn{1}{c}{$\nu_{B}$} & $\chi^{2}/dof$\\
(KeV)       &     &         &             &  \multicolumn{1}{c}{(Hz)} & \\
& & & & &\\   
0.1-0.5 & $2.90\times 10^{-2}$ &$-1.04^{+0.6}_{-0.5}$ & $-2.75^{+0.2}_{-0.4}$ &
$ 5.5^{+ 8}_{ -3} \times 10^{-4}$  & 18.4/8 \\ 
0.5-1.0  &$3.54\times 10^{-3}$ &$-1.27^{+0.5}_{-0.4}$ &$ -2.80^{+0.4}_{-0.7}$&
$12^{+24}_{ -8} \times 10^{-4}$ & 11.1/8 \\
1.0-2.0 & $9.14\times 10^{-2}$ &$ -0.94^{+0.9}_{-*}$ &$ -2.15^{+0.2}_{-0.5}$ &
$ 3.2^{+23}_{ -2} \times 10^{-4}$ & 14.6/8 \\ 
2.0-5.0 & $4.10\times 10^{-4}$ &$-1.45^{+0.4}_{-*}$   &$ -3.28^{+2.2}_{-3.5}$&
$64^{+89}_{-61} \times 10^{-4}$ & 6.5/8 \\ 
5.0-10.0& $4.40\times 10^{-3}$ &$ -1.13^{+0.5}_{-*}$   &$-3.11^{+1.1}_{-1.7}$&
$16^{+22}_{-13} \times 10^{-4}$ & 11.9/8 \\ 
&&&&&\\
0.1-2.0 & $8.80\times 10^{-3}$ & $-1.16^{+0.5}_{-0.4}$ & $-2.86^{+0.3}_{-0.4}$ & 
$ 8.8^{+11}_{ -4} \times 10^{-4}$ & 19.4/8
\\ 
2.0-10.0 & $3.21\times 10^{-4}$ & $-1.47^{+0.3}_{-*}$ & $-4.09^{+*}_{-*}$& 
$71^{+47}_{-53} \times 10^{-4}$ & 6.9/8 \\ 
\end{tabular}
\label{tab:allfree}

$^{*}$ The error is large and no reliable value could be determined.
\end{table*}

\begin{table*}
\centering
\caption {\bf BENDING POWERLAW FITS TO {\em XMM-Newton} PSD OF NGC~4051 WTIH 
FIXED $\alpha_{L}=-1.1$}
\begin{tabular}{lccrr}
Energy Band & Normalisation (A) & $\alpha_{H}$ &\multicolumn{1}{c}{$\nu_{B}$} & $\chi^{2}/dof$\\
(KeV)       &              &              &  \multicolumn{1}{c}{(Hz)} & \\
& & & & \\   
0.1-0.5 &$1.76\times 10^{-2}$  & $-2.77^{+0.2}_{-0.2}$ &$ 5.9^{+ 3}_{-2} 
\times 10^{-4}$ & 18.4/9 \\ 
0.5-1.0 &$1.48\times 10^{-2}$& $-2.69^{+0.2}_{-0.3}$ &$ 8.6^{+ 4}_{ -4} 
\times 10^{-4}$ & 11.5/9\\ 
1.0-2.0 &$2.08 \times 10^{-2}$& $-2.21^{+0.2}_{-0.3}$ &$ 4.5^{+ 5}_{ -3} 
\times 10^{-4}$ & 14.8/9\\ 
2.0-5.0 &$1.08 \times 10^{-2}$& $-2.04^{+0.4}_{-0.6}$& $ 7.4^{+16}_{-7}
\times 10^{-4}$  & 8.8/9\\ 
5.0-10.0&$5.72\times 10^{-3}$& $-3.04^{+0.9}_{-1.2}$ &$15^{+ 9}_{ -8} 
\times 10^{-4}$  & 11.9/9 \\
&&&& \\
0.1-2.0&$1.43 \times 10^{-2}$ & $-2.83^{+0.2}_{-0.2}$ &$ 8.1^{+ 3}_{ -3} 
\times 10^{-4}$  & 19.5/9 \\ 
2.0-10.0 &$9.68\times 10^{-3}$& $-2.03^{+0.4}_{-0.7}$ &$ 8.3^{+19}_{ -8} 
\times 10^{-4}$  &10.5/9 \\ 
\end{tabular}
\label{tab:fixal}
\end{table*}

Fixing $\nu_{B}$ at the average best fit value of
$8\times10^{-4}$Hz, we then see, in Table~\ref{tab:fixalvb} how 
$\alpha_{H}$ decreases steadily with increasing photon energy.
If, instead, we fix $\alpha_{L}$ (at \mbox{-1.1}) and $\alpha_{H}$ (at -3),
$\nu_{B}$ moves to lower frequencies at higher energies.

We note that Papadakis \etal (2002) find a high frequency break in the
ASCA PSD of Akn~564. Vaughan \etal (2003) and Vaughan and Fabian
(2003) find high frequency breaks in the {\em XMM-Newton} PSD of MCG-6-30-15 and
Mkn~766 respectively. These PSDs show the same dependence of
high frequency slope on energy that we find here for NGC~4051
and, where measured, also show no evidence for change of $\nu_{B}$ with energy.
Nandra and Papadakis (2001), in {\it RXTE} observations of NGC7469,
find that the PSD in the $10^{-6}$ to $10^{-3}$Hz range 
is flatter in the 4-10 keV band than in the 2-4 keV band, and
flattens further in the 10-15 keV band. 
The {\it RXTE} PSDs are not of as high quality as the {\it XMM-Newton} PSD
and, particularly, have a gap between  $10^{-4}$ and $10^{-3.5}$Hz and
so it is not possible to fit particularly detailed models.
Nonetheless the results are broadly similar to those reported here,
and elsewhere, with {\it XMM-Newton}.
It is interesting to note that $\alpha_{H}$ 
is now actually better determined in AGN like NGC~4051 and MCG-6-30-15
than in GBHs like Cyg~X-1.

Again, for comparison with other work, we fit the sharply breaking
model to the 0.1-2 keV PSD. Allowing all parameters to be free we find
that $\alpha_{L}=-1.35^{+0.37}_{-0.24}$,
$\alpha_{H}=-2.59^{+0.13}_{-0.15}$,$\nu_{B}=8.6^{+1.5}_{-2.3} \times
10^{-4}$Hz, with a $\chi^{2}/dof$ of 17.8/8. Although $\alpha_{L}$ is
steeper than in the bending powerlaw fit or in the combined {\it RXTE} and
{\em XMM-Newton} fits, and $\alpha_{H}$ is flatter, the value of $\nu_{B}$ is very
similar to that of the bending powerlaw fit, as it the overall 
quality of the fit. Fixing $\alpha_{L}$ at -1.1 has little affect;
$\alpha_{H}$ is unaltered and $\nu_{B}$ changes to $7.4^{+1.5}_{-1.4}
\times 10^{-4}$Hz. The $\chi^{2}/dof$ is 18.9/9.

\subsection{Joint Modelling of {\it RXTE} and 0.1-2 keV {\em XMM-Newton} Data}
\label{sec:xteandsoftxmm}

In our fitting of the \xmm data on its own we have shown that, within
our relatively small errors, $\nu_{B}$ is independent of photon energy
Vaughan \etal (2003) and Vaughan and Fabian (2003) find similar
results from the {\em XMM-Newton} PSD of MCG-6-30-15 and Mkn~766
respectively. We also find, although with larger errors, $\alpha_{L}$
is the same for all energies in NGC~4051.  The same conclusions
regarding both $\nu_{B}$ and $\alpha_{L}$, although with much greater
statistical precision, are true of Cygnus X-1 both in the high
(Cui \etal 1997b; Churazov \etal 2001) and low ( Nowak \etal 1999; Revnivtsev 
\etal 2000) states. Churazov \etal
also show that for Cygnus X-1 the soft disc emission does not
affect significantly the PSD at any energy, except to lower the 
normalisation of the lower energy band PSD by providing a relatively quiescent
component.

The 2-10 keV {\it RXTE} data are largely responsible for defining
$\alpha_{L}$, but the break is more clearly defined in the 0.1-2 keV
{\em XMM-Newton} PSD. Therefore if, based on the observations
described in the previous paragraph, we assume that $\nu_{B}$ and
$\alpha_{L}$ are independent of photon energy in NGC~4051, we can
reduce the uncertainty in all fit parameters by fitting together the
{\it RXTE} and 0.1-2 keV {\em XMM-Newton} data.  As in
Section~\ref{sec:2-10kevpsd} we include, in the fitting process, the
long and medium timescale {\it RXTE} lightcurves together with a 500s
binned 0.1-2 keV {\it XMM-Newton} lightcurve, and allow the
normalisations of all resultant parts of the combined PSD to remain
free. 

\begin{figure*}
\psfig{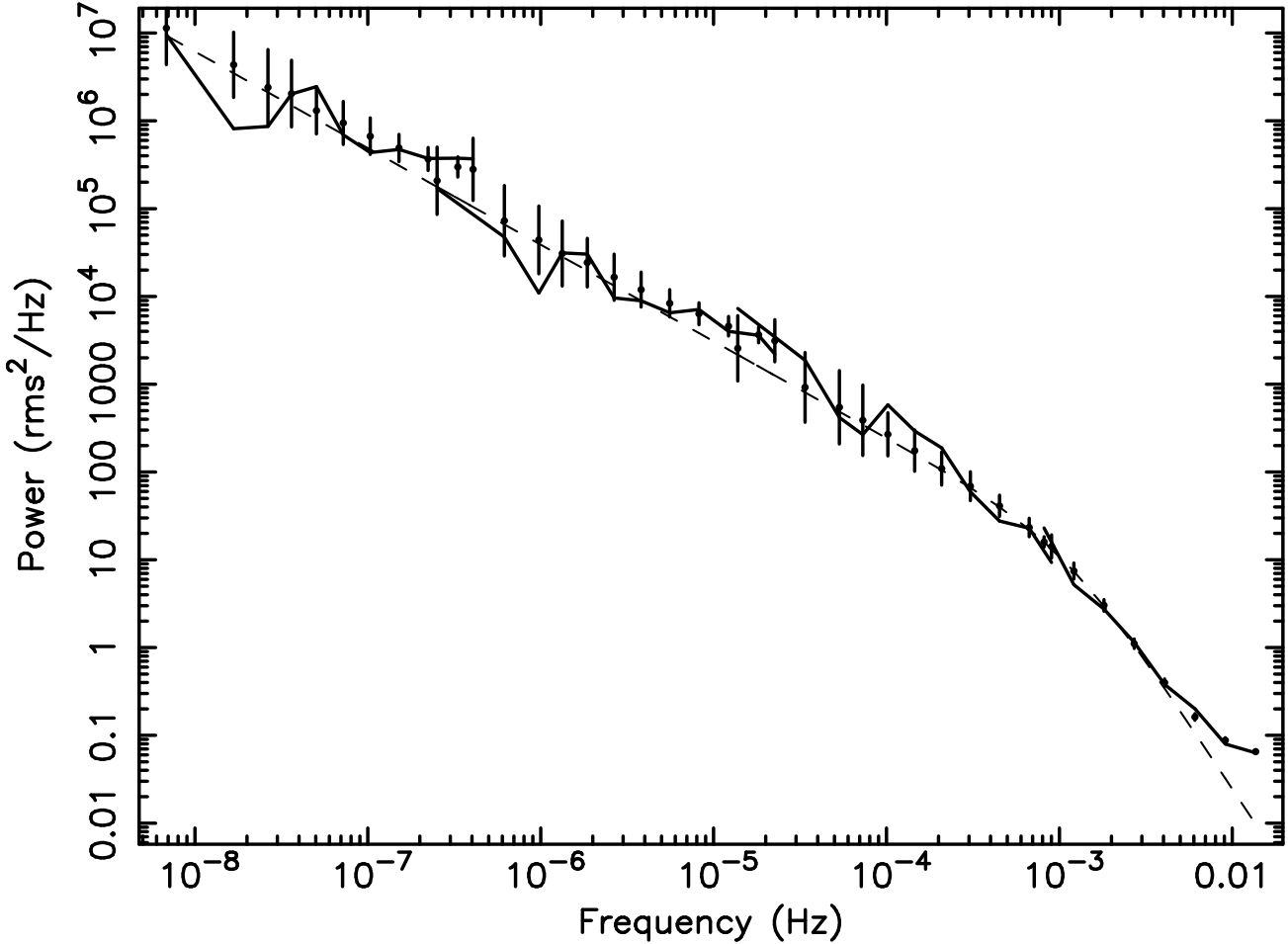}
\caption{Combined {\it RXTE} and {\em XMM-Newton} (0.1-2 keV) PSD. This
PSD is identical to that of Fig.~\ref{fig:xtexmmpsd}
except that the {\em XMM-Newton} data is 0.1-2 keV. Note the very strong
similarity to the high state PSD of Cyg~X-1 (Fig.~\ref{fig:cygpsd}).
}
\label{fig:xtexmm012psd}
\end{figure*}

\begin{figure}
\psfig{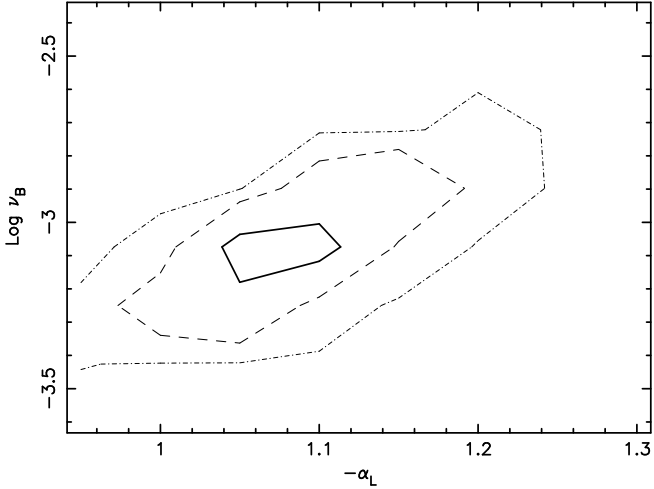}
\caption{68\%, 90\% and 99\% confidence contours for low frequency slope, $\alpha_{L}$
(plotted as $-\alpha_{L}$)
and break frequency, $\nu_{B}$, for bending powerlaw fit to the combined
{\it RXTE} (2-10 keV) and {\em XMM-Newton} (0.1-2 keV) PSD.}
\label{fig:xte012}
\end{figure}

For the very highest frequency part ($>10^{-3}$Hz) of the
combined PSD we incorporate, as discussed in
Section~\ref{sec:analysis}, part of the PSD derived from the {\it
XMM-Newton} 5s resolution 0.1-2 keV lightcurve alone. However in order
to eliminate any spurious breaks or bends caused by joining together
PSDs of different normalisations, the normalisation of the 0.1-2 keV
{\em XMM-Newton} PSD was fixed at that of the 4-10 keV {\em
XMM-Newton} PSD, derived earlier, whose mean energy is the same as
that of the {\it RXTE} observations.  The resulting fit is shown in
Fig.~\ref{fig:xtexmm012psd} and the resulting confidence contours for
$\nu_{B}$ and $\alpha_{L}$ are shown in Fig.~\ref{fig:xte012}. We
obtain quite a good fit (P=53\%) and the results are entirely
consistent with the analysis above using 4-10 keV {\em XMM-Newton}
data. We find that $\nu_{B}=8^{+4}_{-3} \times 10^{-4}$Hz,
$\alpha_{L}=-1.08\pm0.1$ and $\alpha_{H}=-2.9\pm0.25$.

We have also fitted the standard sharply breaking powerlaw to the
combined {\it RXTE} and \xmm 0.1-2 keV data.
The fit  (P=28\%) is worse than the bending 
powerlaw fit although the fit parameters are similar
($\nu_{B}= 6.1 \times 10^{-4}$Hz,  $\alpha_{H}=-2.6$ and
$\alpha_{L}=-1.1$).

\begin{table}
\centering
\caption {\bf BENDING POWERLAW FITS TO {\em XMM-Newton} PSD OF NGC~4051 WTIH FIXED
$\alpha_{L}=-1.1$, $\nu_{B}=8 \times 10^{-4}$Hz}
\begin{tabular}{lccr}
Energy Band & Normalisation (A) &  $\alpha_{H}$ & $\chi^{2}/dof$\\
(KeV)                     &      &   & \\
& & & \\   
0.1-0.5 &$1.34 \times 10^{-2}$  &$-2.91^{+0.12}_{-0.12}$ & 20.0/10  \\
0.5-1.0 &$1.55 \times 10^{-2}$ &$-2.66^{+0.12}_{-0.12}$ & 11.5/10 \\ 
1.0-2.0 & $1.48 \times 10^{-2}$  &$-2.37^{+0.13}_{-0.12}$ & 16.3/10 \\ 
2.0-5.0 &$1.04 \times 10^{-2}$ &$-2.06^{+0.17}_{-0.16}$ & 8.8/10  \\ 
5.0-10.0&$8.04 \times 10^{-3}$  &$-2.31^{+0.47}_{-0.44}$ & 14.2/10  \\ 
&&&\\
0.1-2.0 &$1.45 \times 10^{-2}$ &$-2.83^{+0.10}_{-0.10}$ & 19.5/10  \\ 
2.0-10.0 &$9.84 \times 10^{-3}$ &$-2.03^{+0.15}_{-0.15}$ & 10.5/10  \\ 
\end{tabular}
\label{tab:fixalvb}
\end{table}

\begin{table}
\centering
\caption {\bf BENDING POWERLAW FITS TO {\em XMM-Newton} PSD OF NGC~4051 WTIH FIXED
$\alpha_{L}=-1.1$, $\alpha_{H}=-3$}
\begin{tabular}{lcrr}
Energy Band & Normalisation (A) & \multicolumn{1}{c}{$\nu_{B}$}  & $\chi^{2}/dof$\\
(KeV)       &                   &       \multicolumn{1}{c}{(Hz)}&  \\
& & & \\   
0.1-0.5 &$1.36 \times 10^{-2}$ &$ 8.4^{+ 1.7}_{ -1.5} \times 10^{-4}$  &  21.4/10 \\
0.5-1.0 &$1.14 \times 10^{-2}$  &$12.9^{+ 2.3}_{ -2.0} \times 10^{-4}$  & 14.9/10 \\ 
1.0-2.0 &$9.60 \times 10^{-3}$ & $17.6^{+ 3.4}_{ -2.7} \times 10^{-4}$ & 35.2/10 \\ 
2.0-5.0 &$5.42 \times 10^{-3}$  &$29.8^{+ 8.2}_{ -6.0} \times 10^{-4}$   & 14.7/10 \\ 
5.0-10.0&$5.75 \times 10^{-3}$ &$14.9^{+ 8.1}_{ -4.4} \times 10^{-4}$  & 11.9/10 \\ 
&&&\\
0.1-2.0 &$1.21 \times 10^{-2}$ & $10.3^{+ 1.7}_{ -1.5} \times 10^{-4}$ & 21.1/10  \\ 
2.0-10.0& $5.01 \times 10^{-2}$& $31.8^{+ 7.9}_{ -5.3} \times 10^{-4}$ & 15.4/10 \\ 
\end{tabular}
\label{tab:fixalah}
\end{table}

For the best-fitting bending powerlaw model, then by fixing the high
frequency break at the parameters of the first combined {\it RXTE} and \xmm
0.1-2 keV fit we can search for a second break at lower frequency. We
first constrained the slope below the low frequency break to be zero,
as might be expected in a low-state GBH model. We then find that the
best-fit frequency for the low frequency break is actually the lower
limit that we searched over, ie $3 \times 10^{-9}$Hz, with a 90\%
confidence upper limit of $7 \times 10^{-8}$Hz, ie we do not detect a
second, lower frequency break. We have also allowed both the low
frequency break frequency and the slope below the break to be free
parameters. In that case the best fit for the slope below the low
frequency break turns out to be identical to the slope above that
break, for break frequencies above few $\times10^{-8}$Hz. At lower
frequencies the slope below the break is not well constrained,
although the best fit remains the same as the slope above that break.
So this fit also confirms that we do not find a second, lower
frequency, break.

The separation between the upper 90\% confidence limit on the
frequency of the possible lowest break to a slope of zero and the
lower 90\% confidence limit on the upper break is a factor of $7
\times 10^{3}$. The slope between this range is -1.08.  This highly 
conservative limit on the range is many times larger than the typical
factor $\sim10-30$ between the upper and lower break frequencies in
low state GBHs, at which the slope is also $\sim -1$.  We can
therefore unambiguously rule out any analogy between NGC~4051 and low
state GBHs. However the analogy with high state GBHs is very strong.


\section{The RMS-Flux Relationship}
\label{sec:rms}

Uttley \& M$^{\rm c}$Hardy (2001) showed that the amplitude of
absolute rms variability in small segments of the X-ray lightcurves of
GBHs scaled with the mean flux of the segments, so that as the X-ray
flux increased, so did the variability.  This `rms-flux relation'
appears remarkably linear in GBHs.  Uttley \& M$^{\rm c}$Hardy also
showed that the variability amplitude in {\it RXTE} lightcurves of
Seyfert galaxies increased with X-ray flux, but due to the much slower
variations in AGN lightcurves, could not tell if the relationship was
really linear.  Subsequently, better quality lightcurves have provided
stronger evidence of a linear rms-flux relation in AGN (in Akn~564,
Edelson et al.  2002; and MCG--6-30-15, Vaughan et al. 2003).  With
the excellent quality of the present {\em XMM-Newton} lightcurve, together
with the high variability of this source, we examine the rms-flux
relation of NGC~4051 in similar detail.

To measure the rms-flux relation of the {\it XMM-Newton} 0.1-10~KeV light
curve, we first split the 5-s binned lightcurve into 39 equal-length
segments of 2560~s (512 bin) duration.  For each segment, we measured the
mean flux and the noise-subtracted variance of the segment.  We then
binned the segment variances according to the segment flux, into 4 flux
bins.  Standard errors were estimated from the scatter in individual
segment variance within each flux bin.  The flux-dependent rms is then
obtained by taking the square-root of the variance in each bin
while the errors are propagated through in the normal way.

\begin{figure}
\psfig{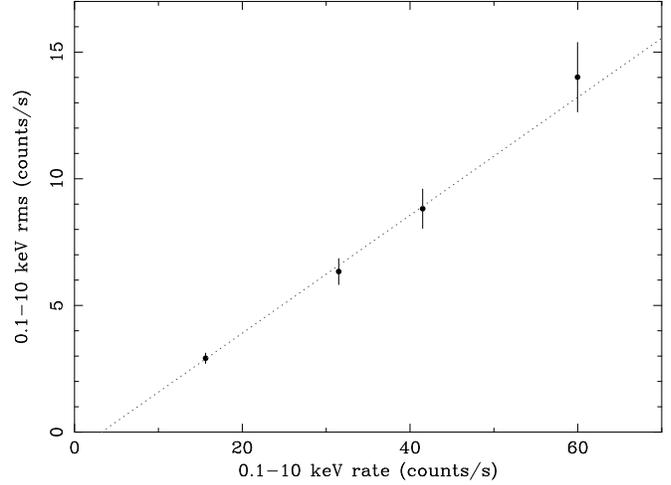}
\caption{rms-flux relationship for NGC~4051 derived from the \xmm
observations in the 0.1-10 keV band. Note the strong linear
relationship and the small offset on the flux axis indicating that
there is a small component of constant rms.
The grey dotted line shows the best-fitting model described 
in the text.}
\label{fig:rmsflux}
\end{figure}

The resulting rms-flux relation for the 0.1-10~keV energy range is
plotted in Fig.~\ref{fig:rmsflux}.  A linear relation with gradient
$k=0.23\pm0.02$ and an flux-axis offset of $C=3.2\pm^{1.5}_{1.7}$
(here 1~$\sigma$ errors) provides a good fit to the data ($\chi^{2}=0.6$
for 2 dof).  The positive offset on the flux axis can be
interpreted as being due to a constant component in the lightcurve
and such offsets are also seen in the rms-flux relations of GBHs (see
Uttley \& M$^{\rm c}$Hardy 2001).  The offset we measure here
corresponds to 8\% of the mean count rate in the 0.1-10~keV band,
compared with the $\sim25$\% offset in GBHs (in bands between 2-13
keV).  To investigate the energy dependence of the rms-flux relation
and so make a better comparison with GBHs, we measured rms-flux
relations in two other bands, 0.1-2~keV and 2-10~keV, obtaining
$k_{0.1-2}=0.24\pm0.02$, $C_{0.1-2}=3.0\pm^{1.2}_{1.4}$ and
$k_{2-10}=0.22\pm0.02$, $C_{2-10}=0.6\pm0.3$.  The measured offsets
correspond to 8\% and 15\% of the mean count rate in each band, for
0.1-2~keV and 2-10~keV band respectively, but the difference between
the two bands is within the errors.  The 90\% upper limit on the
offset in the 2-10~keV band corresponds to 26\% of the mean count
rate, so that we cannot say conclusively that the offset in the
rms-flux relation of NGC~4051 is smaller than that in GBHs, although
we note that a small value is more consistent with the small, or
possibly zero, offset found by Vaughan \etal (2003) in MCG-6-30-15.

\section{Cross Correlation Analysis}
\label{sec:crosscor}

\begin{figure}
\psfig{figure=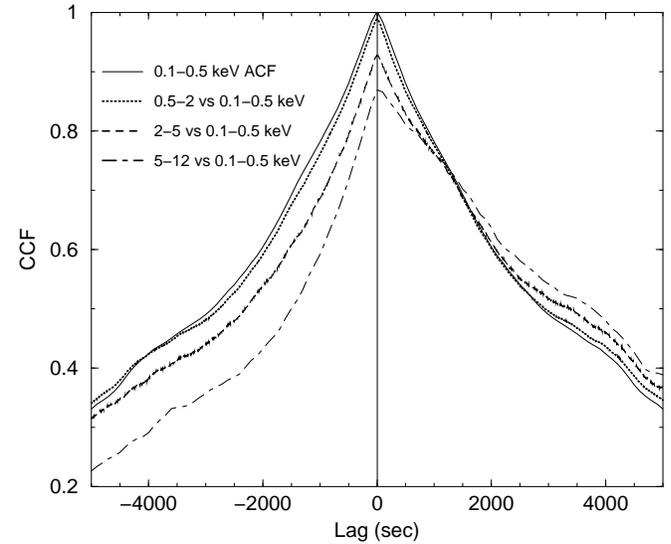,width=3.4in,angle=0}
\caption{The Cross Correlation Function of the $0.1-0.5$ light
curve vs the $0.5-2$ keV (short-dashed line), the $2-5$ keV (long-dashed
line) and the $5-10$ keV (short-long dashed line) lightcurves. The solid
line shows the Auto Correlation Function of the $0.1-0.5$ keV lightcurve.  
The CCFs are all asymmetric towards positive lags, indicative of complex
delays between the hard and soft energy band variations
}
\label{fig:ccf}
\end{figure}

In order to investigate the relationship between the variations in the
different energy bands we first estimated their
Cross-Correlation Function (CCF). We used 100~s binned lightcurves,
and computed the CCF between the $0.1-0.5$ keV lightcurve and the
$0.5-2, 2-5,$ and $5-10$ keV band lightcurves.

The CCF at each lag $k$, $CCF(k)$, was computed as follows:

\[
CCF(k)=\frac{\sum_{t}(x_{soft}(t) -\bar{x}_{soft})(x_{hard}(t+k)-\bar{x}_
{hard})}{N(k)(\sigma^{2}_{soft}\sigma^{2}_{hard})^{1/2}}, \]

\[ k=0,\pm \Delta t,\ldots \].

The summation goes from $t=\Delta t$ to $(N-k)\Delta t$ for $k\geq 0$ and
from $t=(1-k)\Delta t$ to $N\Delta t$ for $k<0$ ($\Delta t = 100$ sec, and
$N$ is the total number of points in the lightcurve), while $N(k)$ is the
number of pairs that contribute to the estimation of the CCF at lag $k$.  
The variances in the above equation are the source variances, i.e. after
correction for the experimental variance. Significant correlation at
positive lags means that the soft band variations are leading the hard
band. All the CCFs were computed up to $k=\pm 10$~ks. 

The resulting CCFs are shown in Fig.~\ref{fig:ccf}. The solid line
curve shows the auto-correlation function (ACF) of the $0.1-0.5$ keV
lightcurve. All the CCFs show a strong peak ($CCF_{max} > 0.8$) at
zero lag. The CCF peak decreases as the energy separation of the light
curves increases. However, the CCFs are not symmetric. This can be
clearly seen when one compares them with the $0.1-0.5$ keV ACF. The
asymmetry is in the sense that the correlation at positive lags is
larger than the correlation at the respective negative lags. 
A broadly similar result is found in {\it RXTE} observations of
NGC7469 by Nandra and Papadakis (2001).
Here in NGC~4051, above a lag $\sim 2$~ks, the $0.1-0.5$ keV lightcurve appears
to correlate better with the $0.5-2$ and $2-5$ keV lightcurves than
with itself. The asymmetry increases with the energy band of the light
curves. The $0.1-0.5$ vs $5-10$ keV CCF shows the strongest asymmetry.

The asymmetry of the CCFs towards positive lags indicates the presence of
complex delays of the higher energy band lightcurves with respect to the
softest band lightcurve. However, the most sensitive tool for the
investigation of energy dependent delays is the phase spectrum, which we
consider in the next section.

\section{Phase Spectrum Analysis}
\label{sec:crosspectrum}

Apart from the CCF, the correlation structure between two lightcurves can
also be characterized by the ``cross spectral density" function (or simply
cross spectrum), which is defined as the Fourier transform of the
cross-covariance function of the two lightcurves. The cross spectrum is a
complex number, and its argument is called the ``phase spectrum". This
function represents the average value of the phase shift between
components with the same frequency in the lightcurves. The corresponding
shift in the time domain is given by the phase shift divided by the
respective frequency, and is usually called the ``time lag" at that
frequency (or corresponding period).

We used 5~s binned lightcurves in the $0.1-0.5$, $0.5-2$, $2-10$, and
$2-5$ keV energy bands, and 100~s binned $0.1-0.5$ and $5-10$ keV light
curves, to calculate the phase spectrum of the higher energy bands with
the respect to the softest band (i.e. $0.1-0.5$ keV). We used
the smaller bin size (5~s) in order to investigate the existence of phase
shifts at the highest possible frequencies. In the case of the $5-10$ keV
lightcurves, due to the low count rate, we used a larger bin size (100~s) in
order to increase the signal to noise ratio. We calculated the real and
imaginary parts of the cross-spectral function, grouped them into period bins of
constant width in log space, and estimated the phase spectrum following
the method described in Papadakis, Nandra and Kazanas (2001).

\begin{figure}
\psfig{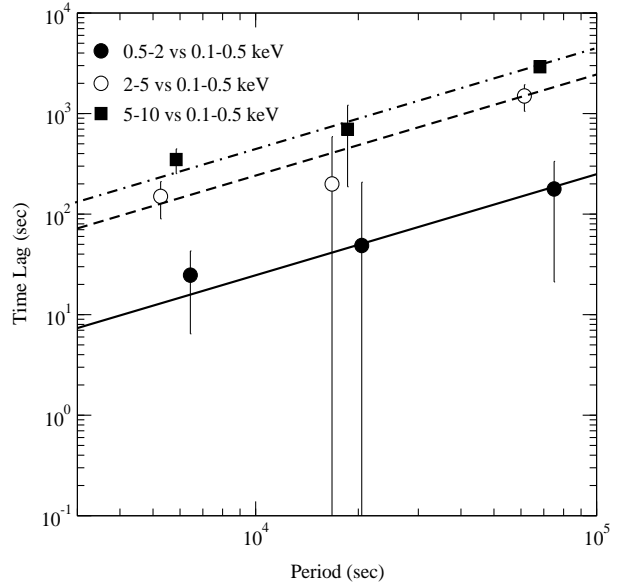}
\caption{Time lag vs. Fourier period for the cross spectrum of
the soft energy band ($0.1-0.5$ keV) vs. the $0.5-2, 2-5, $ and $5-10$
keV bands (filled, open circles and filled squares, respectively).
The sense of the lag is that the hard bands are all delayed with
respect to the soft (0.1-0.5~keV) band. The solid, dashed, and
dot-dashed lines show the best-fitting power law model to the
respective time lag plots, assuming a slope of 1.  Note that the
filled black squares are placed in exactly the correct position but
the open and filled circles have been displaced slightly to lower and
higher periods respectively so that errorbars do not overlap.  For any
given period the lag increases with the energy separation of the
bands.  }
\label{fig:lags}
\end{figure}

Our results are shown in Fig.~\ref{fig:lags}.  In this figure we
show the time lags as function of the Fourier period (i.e.,
$1/\nu$). For clarity reasons we plot only the lags at the three
longest periods, which have the smallest uncertainty. The lags at
these periods show clearly that Fourier components in the higher bands
are delayed with respect to the same components in the $0.1-0.5$ keV
band. The amount of delay does not remain constant but increases with
Fourier period. For example, we observe a $\sim 100$~s delay for the
components with a period of 5~ks, increasing to $\sim 1$~ks delay
for components with a period of 60~ks. This result is similar
to the that seen in other Seyfert galaxies (eg NGC~7469,
Papadakis et al. 2001; MCG 6-30-15, Vaughan et al. 2003).

We fitted a power-law model of the form, $Time Lag \propto \nu^{-s}$, to
the time lags shown in Fig.~\ref{fig:lags}. 
The best fitting slopes were all consistent with the $s=-1$ value that
is commonly found in the time lag plots of GBHs (e.g. Nowak et
al. 1999). Therefore as there are few data points and the slope errors
are large, we fix the slopes in Fig.~\ref{fig:lags} at $s=-1$.  We see
that, at all periods, the lag increases as the separation of the bands
increases, from approximately 0.006\% of the Fourier period for the
$0.1-0.5/0.5-2$~keV lags, to 0.5\% of the Fourier period for the
$0.1-0.5/5-10$~keV lags.

These results are similar to those observed previously in GBHs
(e.g. Nowak \etal  1999) and in MCG-6-30-15 (Vaughan \etal 2003).
The increase in lag with both energy separation of the bands and
with increasing Fourier period also explains the
asymmetric CCFs (Fig.~\ref{fig:ccf}).

\section{Estimation of the Coherence Function}
\label{sec:coherence}

Having computed the complex cross spectra of the various energy bands,
we also estimated their coherence functions.
The value of the coherence function between two lightcurves, at a
certain frequency, may be interpreted as the correlation coefficient
between the Fourier components of the two lightcurves at that
frequency,(Priestley 1981; Vaughan and Nowak 1997).  If the coherence
function is close to unity at all frequencies, we expect to obtain a
close linear relationship between the two lightcurves.

\begin{figure}
\psfig{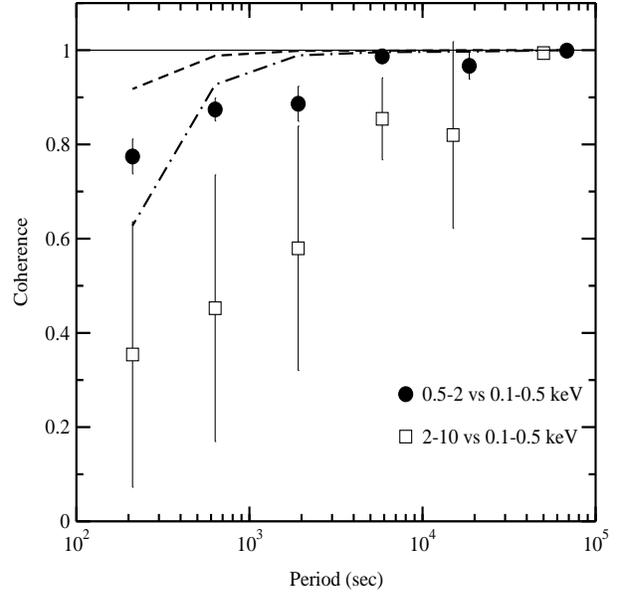}
\caption{Coherence function vs. Fourier period for the soft
energy band ($0.1-0.5$ keV) vs. the $0.5-2 $ and $2-10$ keV bands
(filled circles and open squares, respectively).  Note that, for the
$0.1-0.5$ keV vs $2-10$ keV function, the two open squares at the
longest Fourier period have been displaced to slightly shorter period
to avoid overlapping of their errorbars with the points of the
$0.1-0.5$ keV vs $0.5-2$~keV function, which remain in their correct
positions.  Coherence of unity is indicated by the continuous solid
line.  The dashed and dot-dashed lines show, for the $0.1-0.5$ keV
vs. $0.5-2 $ and $2-10$ keV bands respectively, the 95\% confidence
limit for spurious lack of coherence introduced by the approximation
in the numerical calculation.  We note that, at shorter Fourier
periods, there is a genuine decrease in coherence beyond that
introduced by numerical approximation.  }
\label{fig:coherence}
\end{figure}

Following Papadakis et al. (2001) we computed the coherence functions
of the $0.1-0.5/0.5-2$ keV and the $0.1-0.5/2-10$ keV lightcurves.  We
chose these broad bands to obtain good S/N. Our results are shown in
Fig.~\ref{fig:coherence}. Although the coherence of both functions is
close to unity at the largest periods ($\sim 10^{5}$~s), it decreases
as the Fourier period decreases and as the energy separation between
the bands increases.  This result is similar to that seen in GBHs (eg
Nowak \etal 1999) and recently in other Seyfert galaxies (MCG-6-30-15,
Vaughan \etal 2003; Mkn 766, Vaughan \etal 2003).

The coherence function was corrected for the effects of Poisson noise,
and errors were calculated, using Equation 8 of Vaughan \& Nowak 1997,
which applies in the limit of high PSD power, relative to the Poisson
noise level, and high measured coherence.  However the high power
assumption breaks down at short timescales so, in order to test
whether any of the observed drop in coherence is real, we carried out
Monte Carlo simulations.  We simulated lightcurves with 5~s time
resolution and identical duration to the {\it XMM-Newton} light
curves, choosing underlying PSD models for each band to match the
corresponding best fitting constant $\nu_{B}$ model (see
Section~\ref{sec:highfreqpsd}).  We simulated 200 pairs of light
curves for each of the coherence function measurements, forcing the
lightcurves to be perfectly correlated (unity coherence) by using the
same random number sequence to generate both lightcurves.  We then
approximated the effects of Poisson noise by adding a random Gaussian
deviate to each data point with variance equal to the average squared
error of the lightcurve.  Coherence measurements were made for each of
the 200 lightcurve pairs, and the 95\% lower limit on the coherence in
each bin is plotted in Fig.~\ref{fig:coherence}.  The simulations
clearly show that although the coherence does drop artificially on
short time-scales, the observed drop in coherence is much larger than
expected from this effect alone.  Therefore most of the observed drop
in coherence is intrinsic to the underlying light curves.

\section{Summary of Observational Results}
\label{sec:summary}

In this Section we summarise the main observational results of this paper.
In the next Section (\S~\ref{sec:implications}) we discuss the implications
of these results.

\begin{enumerate}

\item We present $>6.5$ years of well sampled X-ray observations of
NGC~4051 by {\it RXTE}, together with $\sim100$~ks of continuous
observations with {\em XMM-Newton}. The resulting PSD spans over 6.5
decades of frequency and is the best determined AGN PSD yet published.

\item A gently bending powerlaw model is a 
better fit to the combined {\it RXTE} and {\em XMM-Newton} PSD of NGC~4051, although a
sharply broken powerlaw is still an acceptable fit.  The gently
bending model is also a much better fit to the high S/N high state PSD
of Cyg~X-1 than a sharply broken powerlaw. As a gentle bend is also
more plausible physically than a sharp break, we use it as our
standard description of the PSDs of NGC~4051.

\item The combined {\it RXTE} and {\em XMM-Newton} (4-10 keV) PSD has a 
slope, $\alpha_{L}$, of $\sim-1.1$ at low frequencies and steepens,
above a break frequency, $\nu_{B}$, of $8 \times 10^{-4}$Hz, to a
slope, $\alpha_{H}$, of $\sim -2$. There is no evidence of a second
break at lower frequencies.

\item In NGC~4051 $\nu_{B}$ is almost independent of
energy, as in Cyg~X-1 in the high state. 

\item In NGC~4051 $\alpha_{H}$
is steeper at lower energies ($\sim -3$, 0.1-2 keV; $\sim -2$, 2-10
keV) whereas in Cyg~X-1 in the high state $\alpha_{H} \sim -3$
over all of the presently observable spectral range (2-13 keV).

\item NGC~4051 shows the same rms-flux relationship (Uttley and \mch
2001) which is displayed by GBHs, including the presence of small
component of constant rms.

\item As shown by the cross-correlation analysis
(Fig.~\ref{fig:ccf}), variations in different energy bands are very 
well correlated in NGC~4051. However as the
phase-spectral analysis shows
(Fig.~\ref{fig:lags}), there are complex lags
between bands, with the hard photons always lagging the soft photons.
The lag is greatest for variations of large Fourier period, and also
increases as the energy separation between bands increases.

\item The coherence between wavebands 
is very high at long Fourier periods but decreases at shorter periods, and
with increasing separation between wavebands. 
We note that the decrease in coherence becomes noticeable at the
same timescale as the break timescale in the PSD.
\end{enumerate}

\section{Implications}
\label{sec:implications}

\subsection{NGC~4051 and Cygnus X-1: A valid comparison?}

In the soft high state, the energy spectrum of Cygnus X-1 is dominated
by relatively constant thermal emission from the accretion disc below
about 6 keV. Above that energy the spectrum is dominated by a variable
component of hard spectrum.  However in the case of NGC~4051 the
constant soft component accounts for only $\sim10\%$ ($\sim2$ counts
s$^{-1}$) of the average flux in the 0.1-0.5 keV band and probably for
a similar fraction in the 0.5-2 keV band (cf Uttley \etal 2003).
So is it legitimate to compare the PSDs of
Cygnux X-1 and NGC~4051 in bands below 6 keV?

In the case of NGC4051 the spectrum below 2 keV is not a simple
powerlaw. The complex shape during the present observations is
described by Salvi (2003) and Salvi \etal (2003). The rather similar
spectral shape during an X-ray low state Targe of Opportunity (TOO)
observation by {\it XMM-Newton}, triggered as a result of our {\it
RXTE} monitoring, is reported by Uttley \etal (2003). The spectral
shape can be parameterised by a combination of a powerlaw and
Comptonised thermal component. However whatever its exact description,
the whole continuum from 0.1 to 10 keV, apart from a $\sim10\%$
constant component, varies together and is consistent with simple
spectral pivoting about some high energy (Uttley \etal 2003). The TOO
observations are thus entirely consistent with the earlier {\it RXTE}
and {\it EUVE} observations which showed strongly correlated
variability in the X-ray and EUV bands.  Thus, whatever band we sample
in NGC4051, we are sampling overwhelmingly the rapidly variable
component.

In the case of Cygnus X-1 in the soft high state, a detailed
examination of its variability as a function of energy, has been
carried out by Churazov \etal (2001).  They show that the soft
component is relatively quiescent and that the hard component is
responsible for the vast majority of the variability in all bands
which they sample (ie above 2 keV). The only effect of the constant
soft component on the PSD is to reduce its normalisation as one moves
to lower energies where the soft component is more dominant.  Indeed
Churazov \etal show (their Figure 2) a 6-13 keV PSD which is very
similar to that which we show in Fig.~\ref{fig:cygpsd} and they note
that, at lower energies, the normalisation is lower but the shape of
the PSD is the same.
  
Therefore, as far as we are able to determine, over all of the energy
ranges discussed in this paper, it is entirely valid to compare the
shapes of the PSDs of Cygnus X-1 and in NGC~4051. We do, of course,
have to bear in mind the contribution from the constant
components when comparing the normalisations of the PSDs.

\subsection{NGC~4051: High or Low State System?}
The overall long and short timescale PSD of NGC~4051
(Fig.~\ref{fig:xtexmm012psd}) is a very close match to the PSD of
Cyg~X-1 in a high state (Fig.~\ref{fig:cygpsd}), having the same
slopes above and below the break, and with the slope below the break
remaining unchanged for over 4 decades.  It does not look at all like
the PSD of Cyg~X-1 in a low state.  We therefore conclude that
NGC~4051 is the analogue of a GBH in a high state.  This is the first
definite confirmation of an AGN in a high state.

\begin{figure}
\psfig{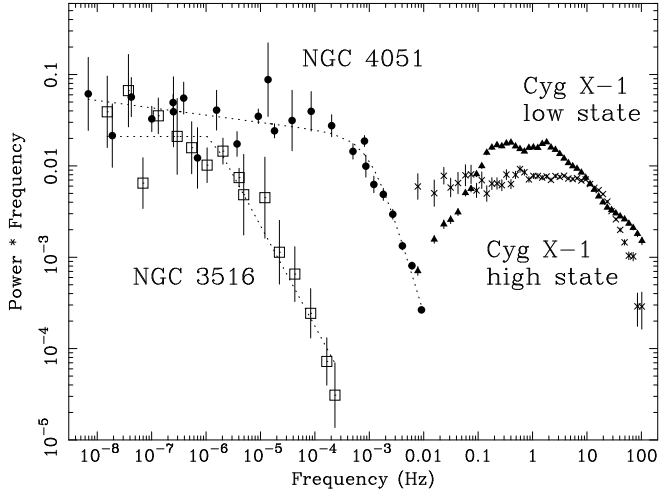}
\caption{Unfolded PSDs of NGC~4051 (filled circles), NGC~3516 (open
squares) and Cyg~X-1 in its low (filled triangles) and high
(asterisks) state. Note that, unlike the other PSDs in this paper, we
show here frequency $\times$ power, not simply power.  }
\label{fig:unfoldedpsd}
\end{figure}

As an alternative representation we plot, in
Fig.~\ref{fig:unfoldedpsd}, the combined {\it RXTE} and {\em
XMM-Newton} (0.1-2 keV) PSD of NGC~4051 in units of frequency $\times$
power ($\nu P_{\nu}$), rather than in units of power.  A flat PSD in
these units has equal power in every decade of frequency and so this
representation is becoming more common as it gives a better idea of
the power distribution. In Fig.~\ref{fig:unfoldedpsd}, we show the PSD
after it has been unfolded from the model fit to the data.  It is the
equivalent of an energy spectrum once it has been deconvolved from the
instrumental response. Thus although the shape of the PSD in
Fig.~\ref{fig:unfoldedpsd} does depend on the assumed model, it is the
closest representation which we can make to the underlying true PSD.
We plot similar PSDs for Cyg~X-1 in its high and low state.  We again
note the similarity between the PSD of NGC~4051 and that of Cyg~X-1 in
its high state. In particular the drop off in power at low frequencies
which is apparent in Cyg~X-1 in its low state is not seen in the PSD
of NGC~4051.

For comparison we also plot the unfolded PSD of the higher mass AGN
NGC~3516 (from Uttley \etal 2002).  The relative scaling of
frequencies with mass is seen quite easily in this figure but we carry
out a more quantitative comparison in Section~\ref{sec:timescales}.

We also note that NGC~4051 has a particularly high variability power
and that Cyg~X-1, in the high state, is somewhat lower. However we
should regard these normalisations with caution as the PSDs have been
normalised by the square of the mean flux. Thus any relatively
constant soft component, such as contributes to the spectrum of Cyg
X-1 in the high state up to $\sim6$keV, raises the mean flux but does
not contribute to the variability.

\subsection{NGC~4051 Black Hole Mass and Eddington Luminosity}
\label{sec:bhmass}

Assuming the bending powerlaw high state model then, assuming a mass
of $10 M_{\odot}$ for the black hole in Cyg~X-1 (Hererro \etal 1995) and
a linear scaling of mass with break timescale, we estimate a black
hole mass for NGC~4051 of $\sim3 ^{+2}_{-1}\times 10^{5}
M_{\odot}$ (assuming a sharply breaking powerlaw implies a mass of
$\sim2 \times 10^{5} M_{\odot}$.)  This (bending powerlaw model)
estimated mass is consistent with the recent reverberation
determination of $5^{+6}_{-3} \times 10^{5} M_{\odot}$ (Shemmer \etal
2003) and the earlier determination of $1.0^{+1}_{-0.5}
\times 10^{6} M_{\odot}$ (Peterson \etal 2000).

For the estimated mass, the Eddington
luminosity, $L_{Edd}$, is $\sim 3 \times 10^{43}$ ergs
s$^{-1}$. Taking an average long term 2-10 keV flux for NGC~4051 of $3
\times 10^{-11}$ \ecs, we calculate a luminosity of $\sim3 \times
10^{41}$ ergs s$^{-1}$. Assuming an X-ray to bolometric conversion
factor of 27  (Elvis \etal 1994; Padovani and Rafanelli 1988),
we deduce that NGC~4051 typically radiates at $\sim30\%$ of $L_{Edd}$.
There is a slight caution that, for NLS1s such as NGC4051, more of
the accretion power may be radiated in the X-rays than for broad
line AGN. However the broad band SED of NGC~4051 presented by Done
\etal (1990) is not too different from the SEDs of typical broad line
QSOs (cf Fig. 10 of Elvis \etal 1994), with the UV/EUV band still
being the dominant band. So although the bolometric correction 
from the X-ray may not be 27 for NGC~4051, it is
still likely to be quite high. Barring better information 
we continue to assume a value of 27 but we note the uncertainty.
Such a high accretion rate compares with typical values of 
$\leq 8\%$ for GBHs when in a low state or
$\gtsim 20\%$
when in the high state (Esin, McClintock and Narayan 1997)
and so strengthens the analogy of NGC~4051 with a high state GBH.
Apart from the NLS1s, most other Seyfert galaxies radiate at 
less than 10\% of their Eddington luminosity (eg Woo and Urry, 2002)
and so might therefore be analogues of low state GBHs. 

\subsection{The Geometry of the Emission Region}

Here we try and place the observational results into some sort of
physical framework. We make the underlying assumption that the X-ray
emission is produced by Comptonisation of low energy seed photons by
energetic electrons in a corona. Before discussing any specific model
we make a number of generic points.

\subsubsection{Physical Interpretation of the Break Timescale}

When considering what physical process the break timescale ($1/8 \times
10^{-4} =1250s$) might correspond to, a useful compilation of dynamical,
thermal, sound-crossing and viscous timescales is given by Treves,
Maraschi and Abramowicz (1988). For a black hole mass of $3 \times 10^{5}
\rm M_{\odot}$, that timescale corresponds to the dynamical timescale,
$T_{dyn}$, at $26R_{G}$ (ie $26GM/c^{2}$). The thermal timescale,
$T_{therm} \simeq T_{dyn}/\alpha$, where $\alpha$ is the viscosity
parameter and the viscous timescale, $T_{visc} \simeq T_{therm} \left(
\frac{R}{H(R)} \right)^{2}$, where $H(R)$ is the scale height of the disc
at radius $R$. The value of $\alpha$ is unknown, but probably
significantly less than unity. If $\alpha < 0.1$, then $T_{therm} > 10
T_{dyn}$, and 1250s could correspond to the thermal time scale at the
innermost part of the disc. In a conventional thin disc, $R>>H(R)$ so
that $T_{visc}$, at a given radius, would be much larger than
$T_{dyn}$.  However in the inner disc $H(R)$ may approach $R$, and
$T_{visc} \sim T_{therm}$. Thus 1250s might correspond to the viscous
or thermal timescales at $\sim$few $R_{G}$.

\subsubsection{PSDs and shot timescale distribution functions} 

Almost any PSD shape can be produced by `shots' with a suitably chosen
distribution of shot timescales (eg Lehto, \mch and Abraham 1991).
Breaks (ie smooth bends) in the PSD are associated with breaks or
cut-offs in the distribution function.  The PSD slope below the break
is related to the slope of the distribution function and the slope
above the break reflects the shape of the individual shots (eg
exponential shots produce a $\nu^{-2}$ PSD). As long as the
distribution function is the same for shots producing photons of
differing energies, we would expect the PSD below the break to be
independent of photon energy. To produce different PSD slopes above
the break at different photon energies, we require either that the
shots have an energy-dependent shape or that the underlying
distribution function does not stop abruptly at some particular
timescale, but that it has a longer tail to
short timescales at higher energies.

\subsubsection{\it Difficulties with Shot Models}

Shot distribution models can produce PSDs which look rather like the
original long and short timescale PSD of NGC~5506 (\mch 1989), and
like the PSDs described here.  However randomly timed shots do not
produce the linear rms-flux relationship which is seen in GBHs and AGN
(Uttley and \mch 2001), and which is followed closely here by
NGC~4051. Some relationship between variations at long and short
timescales is required to produce a linear rms-flux relationship.
A plausible framework for the rms-flux relationship is provided by the
disc model of Lyubarskii (1997) where successively shorter timescales
are associated with annuli successively closer to the black hole.  A
disturbance, perhaps in accretion rate, starts at the outer edge of
the disc and propagates in, thus providing the required link between
variations at long and short timescales.

\subsubsection{ Variations from the seed photons or corona?}

If the corona is uniform, the flatter PSD slope above the break at high
energies cannot be explained if all of the variations come from the
seed photons. In that case, the increased scatterings necessary to
raise seed photons up to the higher X-ray energies would tend to wash
out high frequency variability at high energies, the opposite of what
we see. Thus either the corona is not of uniform temperature, or
fluctuations within the corona must be responsible for introducing at
least some, and possibly most, of the variability at high frequencies.

\subsubsection{The Churazov et al. and Kotov et al. Models} 

By geometrically varying parameters such as the temperature or density
of the emitting region, we can preferentially associate different
photon energies with different radial locations. If we also associate
different variability timescales with different locations we can build
models to attempt to explain the variability of GBHs (eg Hua, Kazanas
and Cui 1999) and AGN (Papadakis \etal 2001).

Recently Churazov \etal (2001), with later enhancements by Kotov \etal
(2001), have built a qualitative geometric model which can explain many
of the observational results presented here (cf Vaughan
\etal 2003).  In the Kotov model a $1/\nu$ powerspectrum of
variations, produced at different radii in the accretion flow
(Lyubarskii 1997), propagates inwards in an optically thin corona,
over the surface of the disc until it hits the X-ray emitting region,
whose emission it modulates. Kotov \etal assume that higher energies
and shorter timescales are associated together at smaller radii.  (One
might, perhaps, associate that region with the region where soft seed
photons from the disc become plentiful.) The break frequency is
associated with a `characteristic timescale' which is, perhaps, the
viscous timescale at the edge of the X-ray emitting region.

{\it Flattening of the PSD at high frequencies:~}
A geometry where shorter timescales and higher energies are associated
together at small radii provides an immediate explanation of the 
flattening of the high, relative to the low, energy PSD at high frequencies.

{\it Decrease of Coherence at Short Fourier Periods:~} If the highest
energies are preferentially produced at the innermost radii, then the
very highest energies see the largest spectrum of variations,
including those at the shortest periods. However if the perturbations
only propagate inwards (as expected if they are perturbations in
accretion rate or, for acoustic perturbations, as a consequence of
increasing viscous timescale with increasing radius) then the lower
energies do not see the shortest period variations and the coherence
decreases as we go to shorter periods and greater separation of the
bands.

{\it Zero lag, asymetric, CCFs:~}
As the bulk of the X-rays will be produced, in all bands,
close to the black hole (eg Fabian \etal 2002 require a highly
centrally peaked emission profile in the disc to explain the shape of
the iron line in MCG-6-30-15) we do not expect large lags between
bands on short Fourier timescales and we expect the CCFs between
bands, although asymetric, to be peaked at zero lag.

{\it Period Dependent Lags:~} However as there will be a larger
component of low energy emission produced further out in the disc
where longer variability timescales apply we expect increased lags at
longer Fourier periods and as the separation of the bands increases.  Any
dispersion, whereby lower frequencies propagate more slowly inwards
towards the main X-ray emitting region, would also lead to period
dependent lags.

{\it Small Differences between NGC~4051 and Cyg~X-1:~} The analogy
between NGC~4051 and Cyg~X-1 in the high state is not absolutely
perfect.  Although we do not sample a great deal of the high state PSD
of Cyg~X-1 above the break, $\alpha_{H}$ appears to be constant with
increasing energy, or maybe even steepens, contrary to the behaviour
of NGC~4051.  The reason for this small difference is not clear,
although we may speculate.  We associate the broad similarities in the
PSDs (ie many decades of slope -1 followed by a
break to a steeper slope at a frequency which is energy-independent)
with broad similarities in the process which gives rise to the
variations.  We have described that process above in terms of
variations propogating inwards over the accretion disc and eventually
modulating the emission from an inner X-ray emitting region. The
smaller differences, ie different behaviour of the high frequency PSD,
may reflect small differences in the detailed structure of the
accretion disc, or emitting corona, which may be caused by differences
in accretion rate or temperature of the disc.

For example we discuss above how a geometry where shorter timescales
and higher energies are associated together at small radii provides an
explanation of the flattening of the high, relative to the low, energy
PSD at high frequencies. However if we can contrive that all energies
are emitted in the same proportion at all radii, then we expect no
change of PSD shape with energy.  Such a scenario might result from a
more conducting disc with less of a radial temperature gradient.

\subsection{Break Timescales for Broad and Narrow Line Seyfert Galaxies}
\label{sec:timescales}

\begin{figure*}
\psfig{figure=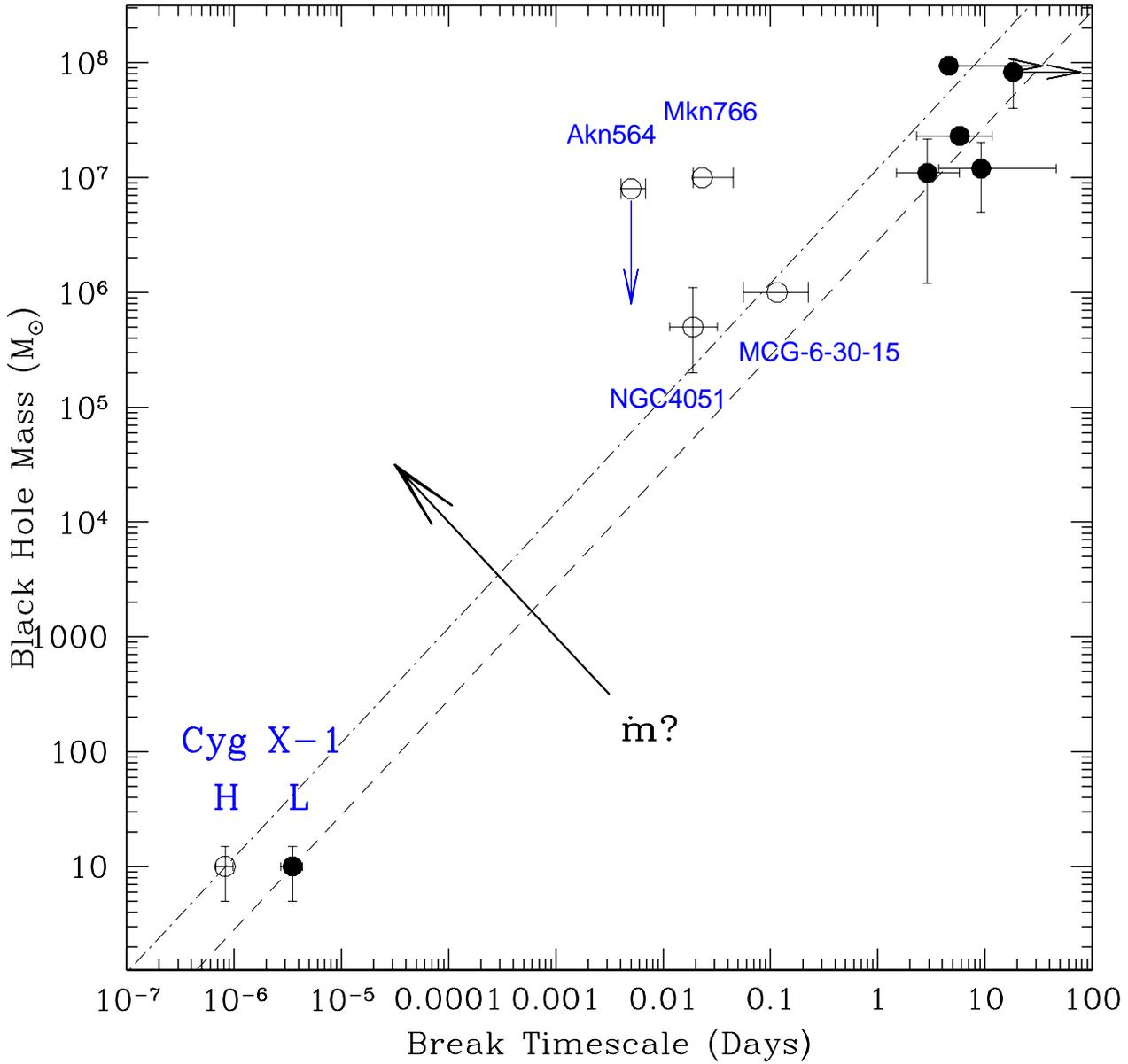,width=6.8in,angle=0}
\caption{Summary of current PSD break timescales for 
AGN of various black hole masses. NLS1s are
shown as open circles and broad line Seyferts are shown as filled
circles.  The high-soft (H) and low-hard (L) states of Cyg~X-1 are
also plotted and lines (dot-dash and long-dash respectively) of slope
1.0 are drawn through those points.  The solid arrow labelled with
$\dot{m}$ indicates the way that the break timescale/mass line may
move with increasing accretion rate.  All break timescales refer to a
break where the PSD slope below the break frequency is $\sim-1$. All
break timescales are derived on the basis of a sharply broken powerlaw
model. See text and Table~\ref{tab:bhdat} for details.}
\label{fig:bh}
\end{figure*}

In Fig.~\ref{fig:bh} we plot the break timescales vs. black hole mass
for Cyg~X-1, in its high-soft (H) and low-hard (L) states, and for AGN
for which break timescales have so far been measured. Narrow line
Seyfert 1 galaxies (NLS1s) are shown as open circles and broad line
Seyferts as filled circles. The data for Fig.~\ref{fig:bh}, together
with relevant references, are given in Table~\ref{tab:bhdat}.  Note
that some of the break timescale in Table~\ref{tab:bhdat} were derived
from a combination of {\it RXTE}  and {\em XMM-Newton}
or {\it Chandra} observations and
so depend to some extent on the assumption that $\nu_{B}$ is
independent of energy. The evidence so far (eg
Sections~\ref{sec:highfreqpsd} and \ref{sec:xteandsoftxmm}) indicates
that this assumption is good but this caveat should be born in mind.

All break timescales refer to the break in the PSD between a low
frequency slope of $\sim-1$ and a high frequency slope of $\sim-2$.
In the context of a low state model, we therefore refer to the higher
of the two PSD break frequencies which, in the case of Cyg~X-1, has
an average value of 3.3 Hz (Uttley \etal 2002).
For Akn~564 we take the break frequency from Papadakis \etal
(2002) rather than that from Markowitz \etal as the Papadakis \etal break
refers to the break between slopes of -1 and -2 whereas the Markowitz
\etal break refers to a break between slopes of 0 and -1.
Note that the masses of MCG-6-30-15, Akn~564 and Mkn~766 are not well
known (cf Wandel 2002).  We therefore do not attempt to place error
bars on these masses, except for the upper limit on the mass of
Akn~564, from Collier \etal (2001).  As the break frequencies derived
from smoothly bending models tend to be higher than those derived from
sharply breaking models (eg 22 and 13.9 Hz respectively for Cyg~X-1,
this paper; 1.3 and 1.0$\times 10^{-4}$Hz for MCG-6-30-15, Vaughan
\etal 2003) we use the break timescales from the sharply breaking
model, as that has been more widely used in the past.  When fitting to
the combined {\it RXTE} and {\em XMM-Newton} 4-10 keV data for NGC~4051 we
find values of 8 and 2.4$\times 10^{-4}$Hz respectively, although the
latter value is not well constrained. When fitting to the
{\it RXTE} and {\em XMM-Newton} 0.1-2 keV data the respective values are 8.0 and
6.1$\times 10^{-4}$Hz. As the ratio of the latter values is similar to
that for MCG-6-30-15 we choose, somewhat arbitrarily, to use the
timescale associated with 6.1$\times 10^{-4}$Hz.

\begin{table}
\begin{centering}
\caption {\bf Black Hole Masses and Break Timescales }
\begin{tabular}{llllc}
\multicolumn{1}{l}{Target} & \multicolumn{1}{c}{PSD Break} &\multicolumn{1}{c}{Ref} 
& \multicolumn{1}{c}{BH Mass} &\multicolumn{1}{c}{Ref} \\
           &        \multicolumn{1}{c}{ Timescale (days)} &     &
\multicolumn{1}{c}{$(10^{7}\, M_{\odot}$)} &  \\
           &                          &     &          &     \\
Fairall~9  & $>18.3$                  & M   & $8.3^{+2.5}_{-4.3}$  & K \\
NGC~5548   & $>4.6$                   & M   & $9.4^{+1.7}_{-1.4}$  & K  \\ 
Akn~564 &$5.0^{+1.8}_{-1.0}\times 10^{-3}$ & P   & $<0.8$               & C \\
NGC~3783   & $2.9^{+2.9}_{-1.4}$      & M   & $1.1^{+1.1}_{-1.0}$& K \\
NGC~3516   & $5.8^{+5.8}_{-3.5}$      & E   & $2.3^{+0.3}_{-0.3}$  & W \\ 
MCG-6-30-15& $1.16^{+1.1}_{-0.6}\times 10^{-1}$  & V,U & $\sim0.1$            & M \\
NGC~4151   & $9.2^{+37}_{-5.5}$       & M   & $1.2^{+0.8}_{-0.7}$& K \\ 
NGC~4051   &$1.89^{+1.3}_{-0.7}\times 10^{-2}$& T & $5^{+6}_{-3}\times 10^{-2}$& S \\ 
Mkn~766    &$2.3^{+2.2}_{-0.4}\times 10^{-2}$ & VF & $\sim1.0$             & W \\
Cyg~X-1 Low &$3.51^{+0.8}_{-0.8}\times 10^{-6}$& N,U &$\sim1\times 10^{-6}$ & H \\
Cyg~X-1 High&$8.26^{+1.5}_{-1.5}\times 10^{-7}$& T&$\sim1\times 10^{-6}$ & H \\
\end{tabular}
\label{tab:bhdat}
\end{centering}

\footnotesize{REFERENCES: \\
C - Collier \etal (2001);
E - Edelson and Nandra (1999);
H - Hererro \etal (1995);
K - Kaspi \etal (2000);
M - Markowitz \etal (2003);
N - Mowak \etal (1999);
P - Papadakis \etal (2002); 
S - Shemmer \etal (2003);
T - This work;
U - Uttley \etal (2002);
V - Vaughan, Fabian and Nandra (2003);
VF - Vaughan and Fabian (2003);
W - Wandell (2002)
}
\end{table}

The first point to note from Fig.~\ref{fig:bh} is that (ignoring the
points without good errorbars) the best fit line (not shown, to avoid
confusion) through all the AGN has a slope flatter than 1,
and does not pass through either of the Cyg~X-1 points.
No line of slope unity passes through all of the AGN and either one
of the Cyg~X-1 points. However
although it would be something of an exaggeration to say that the
timescales associated with broad line AGN scale exactly linearly with
mass with that associated with the low state of Cyg~X-1, the data are
at least consistent with that interpretation. The NLS1s,
however, all lie consistently above the low state line and, indeed,
almost all lie either on or above the high state line, although note
the uncertainties in the masses of 3 of the 4 broad line AGN.
The supposition that break timescales, for a given mass, are shorter
in NLS1s is strongly supported by the observations of
Leighly (1999) who shows that the rms variability of NLS1s
is greater than that of broad line galaxies. This result is easily
explained if the PSDs of broad line
galaxies steepen at lower frequencies than those of NLS1s
and so contain less variability power.

The present data indicate that NGC~4051 is radiating at a substantial
fraction ($\sim30\%$) of $L_{Edd}$ and, assuming a high state model,
Vaughan \etal (2002) draw a similar conclusion regarding MCG-6-30-15.
For Mkn~766 Vaughan and Fabian (2003) note that linear scaling to
the high state model implies radiation at greater than $L_{Edd}$,
and more serious problems would occur with Akn~564 (cf Papadakis \etal 2002).
Scaling the NLS1s to the low state model would make
matters much worse. 

It should also be noted that although we have shown here that NGC~4051
has a PSD which very closely resembles that of a high state system,
we do not yet have such information for Mkn~766 or MCG-6-30-15,
although analysis of most recent {\it RXTE} data should produce the answer
for MCG-6-30-15 (\mch \etal, in preparation).

The PSD of Akn~564 (Papadakis \etal 2002; Pounds \etal 2001) appears to
flatten to a slope of zero at the lowest measured frequencies.  It
could, therefore, be a low state system. However, if so, the mass
implied by scaling its break frequency would be extremely low and
would imply that Akn~564 was greatly exceeding $L_{Edd}$.  An
alternative hypothesis is that Akn~564 is the analogue of a very high
state system (eg GS 1124-68; van der Klis 1995; McClintock and Remillard 2003). 
The PSDs of very high state systems, although of variable shape,
sometime have the same shape as those of low state systems, but 
shifted to higher frequencies. Therefore if we were to add another line into
Fig.~\ref{fig:bh} to represent very high state systems, that
line would lie above the line of the high state systems.

We therefore hypothesise that the black hole mass/break timescale
relationship is not represented by a single relationship for all AGN,
but consists of a family of relationships, moving to shorter
timescales as some underlying parameter (or possibly more than
one parameter) which we have to alter to move
from low to very high state systems changes.

In galactic systems it is generally believed (eg Esin \etal 1997;
McClintock and Remillard 2003, Homan
\etal 2001) that the inner edge of the accretion disc moves closer to 
the black hole as systems change from the low to the high state.  If
the break timescale is associated with the edge of the accretion disc,
and the same physics (eg viscosity) defines the characteristic
timescale in all black hole systems, then movement of the edge of the
disc provides a natural explanation of Fig.~\ref{fig:bh}.  The main
question then would be, what parameter affects the location of the
inner edge of the disc? An obvious possibility would be the accretion
rate, $\dot{m}$ (eg McClintock and
Remillard 2003).  That possibility may apply as long as the location
of the inner edge of the disc is not restricted by the last stable
orbit, ie as long as the inner edge of the optically thick disc
remained outside of the last stable orbit.  However Homan \etal (2001)
suggest that state transitions in the GBH XTE J1550-564 do not depend
greatly on $\dot{m}$.

Alternatively, if the discs in all AGN reach right down to the last
stable orbit, then we have to alter the location of the last stable
orbit and the last stable orbit is closer in for black holes of higher
spin (cf Vaughan and Fabian 2003).  Thus although it is not possible,
on the timescales on which changes are seen, to alter the spin of an
individual GBH (or AGN) in order to change it from a low to a high
state system, it is possible that, for an ensemble of AGN, a range of
spin might provide for a distribution in the mass/break-timescale
plane. In either case, AGN which lie above the dot-dash line in the
Fig.~\ref{fig:bh} should have broader X-ray iron emission lines than
those lying below the long-dash line.  Further 
observations are required to determine break timescales for
a much larger sample of AGN, with a variety of properties, in order to
determine which parameters are most important for determining the
location of AGN in the break timescale/black hole mass plane.

{}

\noindent
{\bf ACKNOWLEDGEMENTS}\\

\noindent
We thank the {\it RXTE} operations team for their great efforts in
scheduling and carrying out the very many observations presented
here. We thank the anonymous referee for a very careful and very rapid
response.  We also thank Chris Done for expert assistance with QDP.
This work was supported by grants from the UK Particle Physics and
Astronomy Research Council (PPARC) including PPA/G/S/2000/00085.
IM$^{\rm c}$H also acknowledges the support of a PPARC Senior Research
Fellowship.

\end{document}